\documentclass[letterpaper,twocolumn,10pt]{article}
\usepackage{usenix}

\usepackage{lipsum}         
\usepackage[utf8]{inputenc} 
\usepackage[T1]{fontenc}    
\usepackage{hyperref}       
\usepackage{url}            
\usepackage{booktabs}       
\usepackage{amsfonts}       
\usepackage{nicefrac}       
\usepackage{microtype}      
\usepackage{xcolor}         
\usepackage{amsmath}

\usepackage{amssymb}
\usepackage{amsthm}
\usepackage{xfrac}
\usepackage{caption}
\usepackage{float}
\usepackage{lscape}
\usepackage{bm}
\usepackage{multirow}
\usepackage{makecell}
\usepackage{algorithm}
\usepackage{algorithmicx}
\usepackage{algpseudocode}
\usepackage{tikz}
\usepackage[all]{xy}
\usetikzlibrary{calc}

\newcommand{\dd}{\mathrm{d}}

\newcommand\<\langle
\renewcommand\>\rangle
\newcommand\ket[1]{\left|#1\right\rangle}

\newcommand\td\tilde
\renewcommand\O{\mathcal O}
\newcommand\A{\mathcal{A}}
\newcommand\lek{_{\le k}}
\newcommand\E{\mathbf{E}}

\renewcommand\l{\ell}
\renewcommand\a{\alpha}
\renewcommand\b{\beta}

\newcommand\sig{\sigma}
\newcommand\Sig{\Sigma}
\newcommand\s{\sig}
\newcommand\eps{\epsilon}
\newcommand\veps{\varepsilon}

\renewcommand\vec[1]{\bm{#1}}
\renewcommand\v[1]{\vec{#1}}

\newcommand\vlj{v_{\l j}}

\DeclareMathOperator*{\tr}{tr}

\DeclareMathOperator*{\rank}{rank}
\DeclareMathOperator*{\laspan}{span}
\DeclareMathOperator*{\argmin}{argmin}

\DeclareMathOperator*{\polylog}{polylog}
\DeclareMathOperator*{\image}{image}
\DeclareMathOperator*{\Prob}{Prob}

\newtheorem{theorem}{Theorem}

\newtheorem{definition}{Definition}

\begin{document}

\date{}

\title{\Large \bf Differential Privacy of Quantum and\\Quantum-Inspired Classical
Recommendation Algorithms}

\author{
{\rm Chenjian Li}\\
Key Laboratory of System Software, CAS \\
University of Chinese Academy of Sciences \\
licj@ios.ac.cn
\and
{\rm Mingsheng Ying}\\
Centre for Quantum Software and Information\\
  University of Technology Sydney \\
Mingsheng.Ying@uts.edu.au
\and
{\rm Ji Guan}\\
Key Laboratory of System Software, CAS\\
guanj@ios.ac.cn
} 


\maketitle

\begin{abstract}
We study the differential privacy (DP) of the quantum recommendation algorithm of Kerenidis--Prakash~\cite{q_recommend_sys} and its quantum-inspired classical counterpart~\cite{ewin_tang_ciq}.
Under standard low-rank and incoherence assumptions on the preference matrix, we show that the randomness already present in the algorithms' measurement/$\l_2$-sampling steps can act as a \emph{privacy-curating mechanism}, yielding $(\veps,\delta)$-DP \emph{without injecting additional DP noise through the interface}.
Concretely, for a system with $m$ users and $n$ items and rank parameter $k$, we prove
$\veps=\O(\sqrt{k/n})$ and $\delta= \O\big(k^2/\min^2\{m,n\}\big)$; in the typical regime $k=\mathrm{polylog}(m,n)$ this simplifies to
$\veps=\tilde \O(1/\sqrt n)$ and $\delta=\tilde \O\big(1/\min^2\{m,n\}\big)$.
Our analysis introduces a perturbation technique for truncated SVD under a single-entry update, which tracks the induced change in the low-rank reconstruction while avoiding unstable singular-vector comparisons.
Finally, we validate the scaling on real-world rating datasets and compare against classical DP recommender baselines.
\end{abstract}

\section{Introduction}

Quantum computation offers a new algorithmic toolbox for exploiting structure in large-scale data.
Beyond canonical speedups such as Shor's factoring algorithm~\cite{shor_algo} and Grover's search algorithm~\cite{grover_algo}, a growing literature investigates quantum acceleration for machine learning and data-analysis tasks~\cite{shor_algo,grover_algo,hhl_algo}.
Among these efforts, \emph{quantum recommendation systems}, initiated by Kerenidis and Prakash~\cite{q_recommend_sys}, have become a landmark example of quantum machine learning: the proposal connects a practically important problem to a concrete quantum subroutine and (under standard data-access assumptions) achieves an exponential runtime improvement over classical approaches at the time of its introduction.
Inspired by the quantum algorithm, Tang later developed an efficient randomized \emph{quantum-inspired} classical algorithm~\cite{ewin_tang_ciq}, which removes the exponential separation while still retaining strong polynomial-time guarantees.

Recommendation systems are now a fundamental component of online services, yet they are inseparable from privacy concerns.
User preferences, even when seemingly innocuous, can enable targeted attacks such as kNN-style attacks~\cite{weinsberg2012blurme} and de-anonymization~\cite{calandrino2011you, narayanan2006deanonymize}.
This has motivated extensive work on designing differentially private (DP) recommendation algorithms~\cite{mcsherry2009Differentially,yang2017Privacy_JL2,berlioz2015Applying,friedman2016Differential}.
A common theme in this line of work is to achieve privacy by injecting carefully calibrated randomness (e.g., Laplace/Gaussian noise) into intermediate statistics or learning updates, often trading off personalization quality for formal privacy guarantees.

Just like their classical counterparts, quantum algorithms can also be susceptible to privacy leakage when driven by sensitive data.
While the computational complexity of quantum recommendation algorithms has been extensively studied, to the best of our knowledge, there has been no prior work that quantitatively characterizes their \emph{differential privacy} guarantees.
At the same time, quantum algorithms are intrinsically randomized due to sampling and measurement.
This raises a natural question:
\begin{quote}
\emph{Can the intrinsic randomness already present in quantum (and quantum-inspired) recommendation algorithms itself act as a privacy-curating mechanism, without injecting additional DP noise through the recommendation interface?}
\end{quote}
We answer this question by analyzing the privacy properties of both the quantum recommendation algorithm~\cite{q_recommend_sys} and the quantum-inspired classical algorithm~\cite{ewin_tang_ciq} within the standard differential privacy framework~\cite{dwork_algodpbook}.
Informally, differential privacy requires that changing the input database in one record (or one user's contribution, depending on the neighboring relation) does not significantly change the distribution of any output event, quantified by a multiplicative parameter $\veps$ and an additive parameter $\delta$.
Our main finding is that, under standard structural assumptions on the underlying preference matrix (low-rank and incoherence), \emph{both} algorithms satisfy DP guarantees that improve with the problem dimension, leveraging only their \emph{native} sampling randomness (quantum measurement and $\l_2$-sampling), rather than externally injected DP noise.

\begin{theorem}[Main Result Preview, Informal]
\label{thm:intro-informal}
Consider a recommendation system with $m$ users and $n$ products.
Under standard low-rank and incoherence assumptions, the quantum recommendation algorithm $\A_{\rm RQ}^k$~\cite{q_recommend_sys} and the quantum-inspired classical recommendation algorithm $\A_{\rm RC}^k$~\cite{ewin_tang_ciq} satisfy $(\veps,\delta)$-DP with
\begin{equation}
(\veps,\delta)
=
\Big(\O\big(\sqrt{k/n}\big),\ \O\big(k^2/{\min}^2\{m,n\}\big)\Big),
\end{equation}
where $k$ is the rank parameter.
In particular, in the
typical regime $k=\polylog\,(m,n)$, this simplifies to
$\big(\td\O(1/\sqrt n),\ \td\O(1/\min^2\{m,n\})\big)$-DP, where $\td\O(\cdot)$ hides $\polylog(m,n)$ factors.
\end{theorem}

\paragraph{Technical challenge and our technique.}
Both the quantum and quantum-inspired algorithms are based on truncated singular value decomposition (SVD) and low-rank matrix reconstruction.
A core obstacle is that SVD captures the global properties of the original matrix: even a single-entry change of the input matrix may cause change in all singular values and vectors in a non-trivial manner. This numerical complexity has also hindered previous attempts to develop tight privacy analyses for matrix-mactorization-based recommendation systems~\cite{friedman2016Differential}.
To overcome the challenge, we introduce a perturbation method tailored to truncated SVD under a single-entry (rank-one) update. 
Our method tracks how the \emph{low-rank reconstruction} changes (in the quantities that determine the output distribution) by isolating its leading-order effect and exploiting the rank-one property.

\paragraph{Comparison to classical DP recommendation systems.}
Finally, we compare the quantum and quantum-inspired recommendation algorithms against classical DP recommendation approaches.
The key distinction is conceptual: classical DP recommenders typically enforce privacy by \emph{explicit} noise injection, while our guarantees arise from the \emph{intrinsic} sampling randomness already used by the algorithms.
This yields a clean privacy--utility comparison under the same neighbouring relation: to match the same $\veps$ level, classical baselines generally require substantially larger injected noise as the problem size grows, whereas the quantum/quantum-inspired algorithms do not introduce additional DP noise beyond their native sampling steps. Accordingly, our results suggest that, under the same interface and neighbouring relation, quantum and quantum-inspired recommenders can \emph{avoid the additional utility degradation caused by explicit DP noise}, and thus offer a better privacy--utility tradeoff, especially in large-scale regimes.

{\vskip 3pt}
In summary, the main contributions of our paper are as follows:
\begin{enumerate}
  \item \textbf{First DP characterization for quantum and quantum-inspired recommendation algorithms.}
  We provide the first differential privacy analysis of the quantum recommendation algorithm~\cite{q_recommend_sys} and the quantum-inspired classical recommendation algorithm~\cite{ewin_tang_ciq} under standard low-rank and incoherence assumptions.
  \item \textbf{Dimension-improving $(\veps,\delta)$ bounds from intrinsic randomness.}
  We show that both algorithms satisfy
  $(\veps,\delta)=\big(\O(\sqrt{k/n}),\O(k^2/\min^2\{m,n\})\big)$-DP ,
  leveraging only the algorithms' native sampling randomness rather than additional DP noise injection.
  \item \textbf{A truncated-SVD perturbation method under single-entry updates.}
  We introduce a new perturbation technique for truncated SVD under rank-one matrix perturbations, enabling stable control of low-rank reconstructions needed for privacy analysis.
\end{enumerate}

\paragraph{Organization of the paper.}
In Sec.~\ref{sec2:preliminaries}, we introduce the problem setting and its background, including quantum computation, the recommendation problem and differential privacy. In Sec.~\ref{sec3:algorithms}, we briefly review the quantum and quantum-inspired recommendation algorithms and our simplified models for them. In Sec.~\ref{sec4:threat_model}, we articulate the threat model used in our privacy analysis.
In Sec.~\ref{sec5:method}, we present the technical tools used in our analysis.
Finally in Sec.~\ref{sec6:dp_res}, we state and prove our main DP results and discuss their implications.
After that, in Sec.~\ref{sec7:exp} we evaluate our findings on real-world datasets and compare against prior differentially private recommendation systems.

\section{Preliminaries}\label{sec2:preliminaries}

\subsection{Quantum Computation}\label{sec:intro_qc}
  In this subsection we provide a basic introduction to quantum computation \cite{nielsen_chuang_qcqi}. For readers seeking a more comprehensive overview, we recommend \textit{Quantum Computation and Quantum Information} by Nielsen and Chuang \cite{nielsen_chuang_qcqi} as an authoritative and comprehensive reference.

  Classical data are represented in bits, whose value is \textit{either} 0 or 1. In contrast, a quantum bit (qubit for short), which is a quantum system with two levels being $\ket{0}$ and $\ket{1}$, can be in arbitrary \textit{superposition} of both states. Specifically, the state of a qubit can be expressed as  $\a\ket{0}+\b\ket{1}$, where $\a,\b$ are complex numbers satisfying the normalization relation $|\a|^2+|\b|^2=1$. It is only upon subsequent measurement that the qubit collapses to either $\ket 0$ or $\ket 1$, with probability $|\a|^2$ and $|\b|^2$ respectively. The coefficients $\a,\b$ are called probability amplitudes. Furthermore, an $n$-qubit system can be in a state
  \begin{equation}
  \ket{\psi}
  =\sum_{\v {i}\in\{0,1\}^n}\a_{\v{i}}\ket{\v{i}},
  \qquad
  \sum_{\v{i}\in\{0,1\}^n}|\a_{\v{i}}|^2=1,
  \end{equation}
  where $\{\ket{\v{i}}:\v{i}\in\{0,1\}^n\}$ is called the \emph{computational basis}, and we have abbreviated the indices into a vector $\v{i}=(i_1,i_2,\ldots,i_n)$.  A computation on such quantum state simultaneously applies to all $N=2^n$ computational basis, achieving an exponential speedup, which is believed to be the source of the computational advantage of quantum computation.

  After being measured, the multi-qubit quantum state $\sum_{\v i\in\{0,1\}^n}\a_{\v i}|\v i\rangle$ collapses to one of the computational basis states, say $\ket{\v j}$, with probability $\Pr(\v j)=|\a_{\v j}|^2$,
  respectively. Note that the quantum measurement is a probabilistic process, and can be naturally used as a sampling mechanism. In particular, quantum measurement in the computational basis can be modelled as a $\l^2$-norm sampling process:
  \begin{definition}[$\l^2$-norm sampling]
    Given a vector $\v v\in\mathbb C^N$, $\l^2$-norm sampling is to pick a number $i\in\{1,2,...,N\}$ with probability
    \begin{equation}
      \Prob_{j\sim \l^2(\v v)}(j)=\frac{|v_j|^2}{\sum_{i=1}^N|v_i|^2}.
    \end{equation}
  \end{definition}

  In quantum machine learning algorithms, data must be encoded into quantum states before any computations can be applied. The most commonly used encoding scheme is the amplitude encoding, which represents data in the amplitudes of a quantum state:
  \begin{definition}[Amplitude Encoding]
    The vector state $\ket{\v v}$ for $\v v\in\mathbb R^N$ is defined as 
    \begin{equation}
        |\v v\>:=\frac{1}{|\v v|}\sum_{\v i=1}^N v_{\v i}\ket{\v i}.
    \end{equation}
  \end{definition}

\subsection{Linear Algebra}

  Singular value decomposition (SVD) plays a central role in the quantum recommendation algorithm and our analysis on it. The singular value decomposition states that any rectangular matrix $A\in\mathbb C^{m\times n}$ can be decomposed as $A=U\Sig V^\dag$
  where $U,V$ are $m\times m$ and  $n\times n$ unitary matrices respectively, and $\Sig$ is a rectangular diagonal matrix with $\Sig_{\l\l}=\sig_\l$. The diagonal entries $\sig_\l$ are called singular values, satisfying $\sig_1\ge \sig_2\cdots\sig_{\min\{m,n\}}\ge 0$. When $A$ is a real matrix, $U$ and $V$ reduces to real orthogonal matrices, and the Hermitian conjugate $V^\dag$ reduces to the real transpose $V^T$.

  While the matrix form $A=U\Sig V^\dag$ of SVD is widely used, in our context it is more convenient to express the SVD in its vector form:
  \begin{equation}
    A=\sum_{\l=1}^{\min\{m,n\}}\sig_\l \v u_\l \v v_\l^\dag
  \end{equation}
  where $\v u_\l$ and $\v v_\l$ are the $\l$-th left and right singular vector of $A$, respectively, and they happens to be the $\l$-th column vector of $U$ and $V$.
  When it is clear from the context, we will abbreviate $\sum_{\l=1}^{\min\{m,n\}}$ as $\sum_\l$. Furthermore, we define the low-rank approximation of $A$ via SVD as 
  \begin{equation}
      A_{\le k}:=\sum_{\l=1}^k\sig_\l\v u_\l\v v_\l^\dag.
  \end{equation}

  SVD is usually connected with the Frobenius norm of matrix. The Frobenius norm of a matrix $A$ is defined as $\|A\|_F=\sqrt{\sum_{i,j}|A|_{ij}^2}$. Using the identity $\|A\|_F^2=\tr(A^\dag A)$, we see that $\|A\|_F^2=\sum_\l \sig_\l^2$. As Frobenius norm is compliant with the Euclidean norm of vectors, we will abbreviate $\|A\|_F$ as $|A|$ for simplicity.




  We will use $V_i$ to denote the $i$-th row of a matrix $V$ and $V_{ij}$ to denote the $(i,j)$-th entry of $V$. We will also use $v_i$ to denote the $i$-th component of vector $\v v$. 

  For a quantity $Q$, we usually use $\tilde Q$ to denote an approximation of $Q$. When quantity $Q_1$ is far greater (or far less) than $Q_2$, we denote $Q_1\gg Q_2$ (or $Q_1\ll Q_2$).

\subsection{Recommendation Problem}\label{sec:recommen_prob}
  With the rapid expansion of online platforms like Amazon and Netflix in recent years, one major challenge they face is recommending new products that align with users' interests based on existing data. This gives rise to the recommendation problem, which requires using machine learning approaches to find patterns in user behaviour and predict products that users are likely to enjoy.

  The recommendation problem can be modelled as follows. Suppose there are $m$ users and $n$ products, then the collected data is commonly modelled as a matrix $T$, with the value of $T_{ij}$ representing user $i$'s rating on product $j$. Without loss of generality, we would restrict $T_{ij}\in\{0,1\}$ in our analysis for simplicity, following previous works on the quantum and quantum-inspired classical recommendation algorithms \cite{q_recommend_sys,ewin_tang_ciq}. Specifically, the entry of recommendation database matrix $T$ is defined as:
  \begin{equation}
  T_{ij}=\left\{\begin{array}{cl}
      1 & \text{user } i \text{ likes product } j\\
      0 & \text{user } i \text{ dislikes product } j \text{ or no data}
  \end{array}\right..
  \end{equation}

  Apparently it is impossible to collect all users' preference on all products, so the matrix $T$ is usually considered a subsampled matrix of the \emph{true preference matrix} $P$, which is inaccessible to the algorithms. The recommendation problem is then reduced to recovering the underlying true preferences $P$ from database $T$ and finding a high value entry in the recovered matrix. The recovering process is called \emph{matrix reconstruction}, or \emph{matrix completion}, which is usually done by pattern recognition and machine learning methods such as collaborative filtering or low-rank projection. 

  The matrix completion problem is an underdetermined problem without additional structures. To make the reconstruction process feasible, two key assumptions underlie any matrix-completion-based recommendation system.
  
  \paragraph{The Low Rank Assumption.} \setcounter{assumption}{1} In the recommendation problem, it is usually assumed that the rank of $P$ is much lower than $m$ and $n$, typically on the order of $\polylog(m,n)$. The assumption implies that there are identifiable and learnable patterns within the data, making the recommendation task possible. The intuition behind the assumption is that most users belong to a small number of groups, and users in the same group tend to have similar preferences. This assumption has been extensively studied \cite{rich1979user,azar2001spectral,tao2010matrixcompletion} and is widely accepted within the recommendation research community \cite{q_recommend_sys,ewin_tang_ciq,candes2012exact,koren2009matrix,jain2013provable}.

  The low rank assumption makes it possible to reconstruct $P$ from $T$, which would otherwise be an ill-conditioned problem.
  The following theorem gives an effective method for reconstructing $P$ based on SVD:

  \begin{theorem}[Eckart--Young \cite{eckart-young}, Best low-rank approximation]
    Suppose $A\in\mathbb C^{m\times n}$ is a complex matrix and its SVD is $A=\sum_\l \sig_\l \v u_\l \v v_\l^\dag$. Then the best low-rank approximation of $A$ is given by:
    \begin{equation}
      \argmin_{\rank(\td A)\le k}|A-\td A|=\sum_{\l\le k} \sig_\l \v u_\l \v v_\l^\dag=A_{\le k}.
    \end{equation}
  \end{theorem}
  The Eckart-Young theorem actually suggests a practical recommendation algorithm via SVD: first, reconstruct the preference matrix as $\tilde P=T_{\le k}$, where $k$ is a rank cutoff parameter tuned by algorithm practitioners. As shown by Achlioptas and McSherry \cite{achlioptasFastComputationLowrank2007}, the low-rank components $T_{\le k}$ closely approximate original matrix $P$ with high probability. Consequently, then sampling a high value entry from $\td P$ would yield reasonably good recommendations. This is shown in Fig.~\ref{fig:Eckart_Young}.

  \begin{figure}[H]
  \centering
  \begin{tikzpicture}[node distance=2.5cm, auto]
    \node (P) at (0,0) {$P$};
    \node (T) at (0,-1) {$T$};
    \node (Tk) at (2,-1) {$T_{\leq k}$};
    \node (Ptilde) at (2,0) {$\tilde{P}$};

    \draw[->] (P) to node[left] {subsample} (T);
    \draw[->] (T) to node[below] {SVD} (Tk);
    \draw[->] (Tk) to (Ptilde);

    \node [above] at ($(P)!.5!(Ptilde)$) {\cite{achlioptasFastComputationLowrank2007}};
    \node at ($(P)!.5!(Ptilde)$) {$\approx$};

  \end{tikzpicture}
  \caption{An illustration of the low-rank recommendation algorithm process implied by the Eckart-Young theorem.}
    \label{fig:Eckart_Young}
  \end{figure}

  \paragraph{The Incoherence Assumption.} Aside from the low-rank requirement, successful matrix completion also relies on a structural regularity of the singular vectors, commonly phrased as an incoherence (or delocalization) condition, which roughly means the singular vectors do not have any large coordinates. 
  
  Intuitively, if some singular vectors are extremely sparse and are concentrated on few coordinates, then the outlier entries cannot be reliably inferred from the rest of the matrix and matrix completion would be ineffective and infeasible, as argued by Tao \cite{tao2010matrixcompletion}. Therefore, effective matrix completion from a small fraction of observed entries requires an incoherence assumption on the matrix that “spreads” the singular vectors across coordinates, rather than concentrating most of their mass on just a few entries.

  Fortunately, low coherence is widely observed in large real-world and random matrices. Random matrices have been proved to exhibit low coherence \cite{rudelson2017delocalization}, as well as many real-world matrices \cite{hardt2013beyond}. Indeed, many recent results in matrix completion, Robust Principal Component Analysis and Low-rank approximation rely crucially on the assumption that the input matrix has low coherence \cite{hardt2013beyond}.

  \begin{figure}
    \centering
    \includegraphics[width=.48\textwidth]{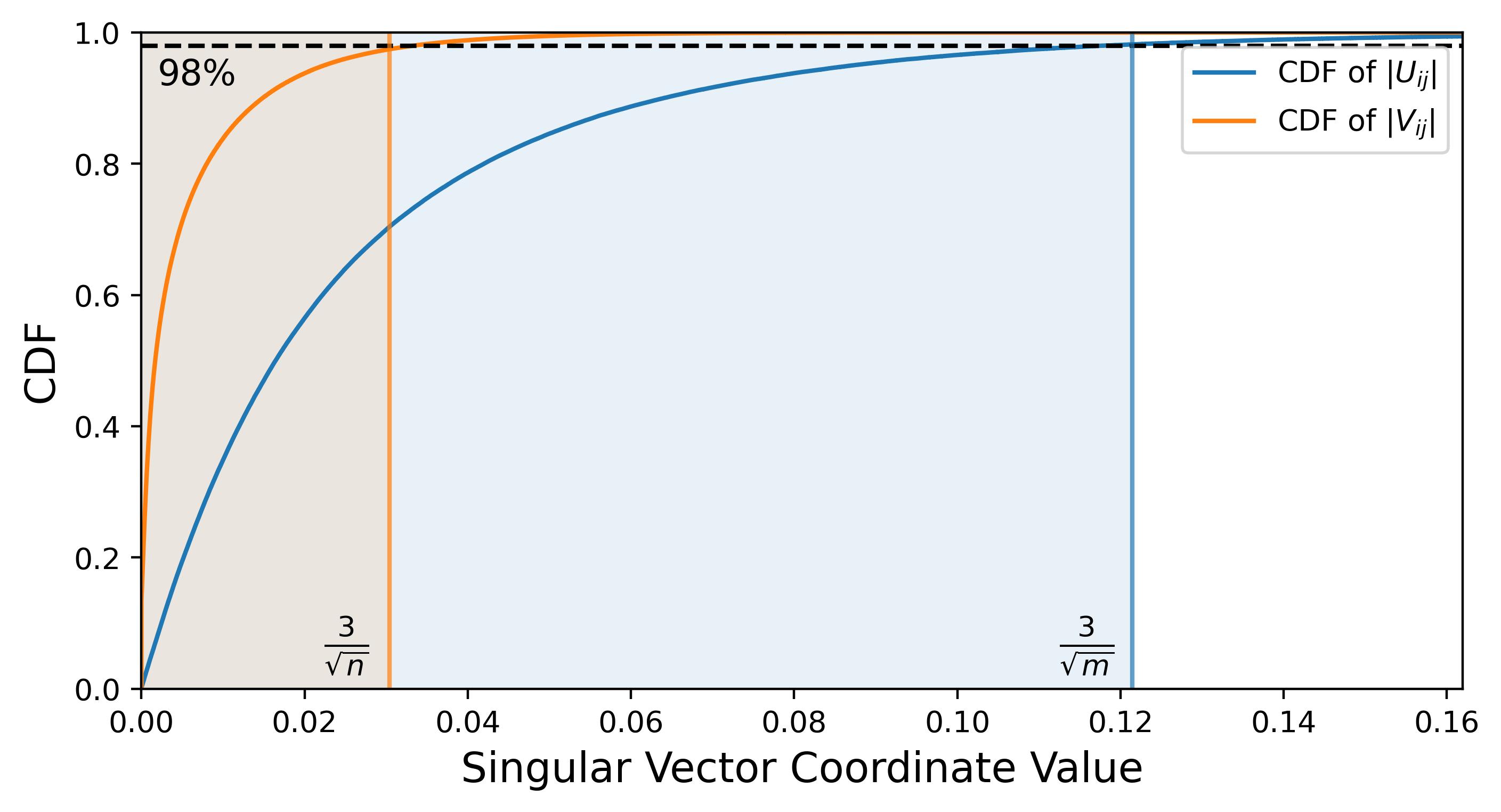}
    \caption{Cumulative distribution function (CDF) of all singular vector coordinates $|U_{ij}|$ and $|V_{ij}|$ on MovieLens-Small \cite{dataset_movielens} dataset. 98\% of left and right singular vector coordinates falls within $\frac{3}{\sqrt{m}}, \frac{3}{\sqrt{n}}$ range respectively, demonstrating the incoherence of the dataset. Experiments on other versions of MovieLens and other datasets show similar and consistent trend.}
    \label{fig:uv_stat}
  \end{figure}

  Formally, Cand\`{e}s and Tao \cite{candes2010power} show that almost all $m\times n$ matrices obeys the the incoherence property
  \begin{equation}
      ||\v u_\l||_\infty\le \sqrt{\mu/m},\quad ||\v v_\l||_\infty\le \sqrt{\mu/n},
  \end{equation}
  where $\mu=\O(1)$ is a constant that characterize the coherence of the matrix. That to say, coordinates of incoherent left and right singular vectors are in the order of $\O(\sfrac{1}{\sqrt m})$ and $\O(\sfrac{1}{\sqrt{n}})$, respectively. In practice, it can be observed via numerical experiment that 98\% of the singular vector coordinate absolute values fall within $\frac{3}{\sqrt{m}}, \frac{3}{\sqrt{n}}$ range for the famous MovieLens dataset  \cite{dataset_movielens} (see Fig. \ref{fig:uv_stat}).

  \subsection{Differential Privacy}
  
  
  Differential privacy (DP) is a provable privacy model based on the principle that the output of a query or computation should not allow inference about any particular record in the input dataset. This is achieved by requiring that the probability of any outcome of a computation is insensitive to small input changes.
  In practices, databases are usually modelled as a collection of records, and neighbouring databases are those that differ in only one entry, denoted as $D\sim D'$.
  Formally, we have the following definition of differential privacy:
  \begin{definition}[Differential Privacy \cite{dwork_algodpbook}]
    A randomized algorithm $\A$ is $(\eps,\delta)$-differentially private if for all neighbouring databases $D\sim D'$, and for all subset $S\subseteq\image \A$,
    \begin{equation}
      \Pr\big(\A(D)\in S\big)\le e^\veps\Pr\big(\A(D')\in S\big)+\delta.
    \end{equation}
  \end{definition}

  In DP terminology, such randomized algorithms are known as \emph{curating mechanisms}. Most of these mechanisms add noise to the data to obscure them to achieve differential privacy, such as the Laplace mechanism \cite{dwork_algodpbook}. Low levels of $\veps$ and $\delta$ corresponds to high degree of privacy. Depending on different setting or user requirements, value ranges like $\veps\le \ln 2$ to $\veps\le 3$ are considered as providing acceptable levels of privacy in the literature.

  \paragraph{DP for recommendation problem.} Though preference data may seem insensitive at first glance, numerous attacks and research \cite{weinsberg2012blurme, calandrino2011you, mcsherry2009Differentially, narayanan2006deanonymize} have been proposed to show that it is necessary to develop differentially private recommendation systems. For example, once de-anonymized \cite{narayanan2006deanonymize}, they may be related to sensitive personal data such as IP address, timing information.

  For the recommendation problem, a record is usually a tuple in the form of (user $i$, product $j$, rating $T_{ij}$), corresponding to an entry in the preference database matrix $T$. Two neighbouring relations are commonly studied in the recommendation literature: While (i) some research \cite{mcsherry2009Differentially, shin2018privacy} focus on user-level DP, where all records corresponding to a user is added or removed between neighbouring databases, (ii) most research \cite{mcsherry2009Differentially, yang2017Privacy_JL2, berlioz2015Applying, friedman2016Differential} focus on record-level DP, where neighbouring databases differ by a single rating record.
  
  In our work, we focus on record-level DP of recommendation system, where neighbouring databases are represented as binary matrices $T,T'$ that differ at exactly one entry $(p,q)$-th entry. Equivalently, one may write $T'=T\oplus \v e_p \v e_q^\dag$, where $\oplus$ is the addition modulo 2. 
  Our goal is to bound $(\veps,\delta)$ such that for all neighbourint $T\sim T'$, the following inequality holds:
  \begin{equation}\begin{aligned}
    \Pr\big(\A_{\rm RQ}^k(T,i)=j\big)
    \le
    e^\eps \Pr\big(\A_{\rm RQ}^k(T',i)=j\big) + \delta\\
    \forall\  \text{user }i,\forall\  \text{product }j .
  \end{aligned}\label{eq:dp_prop}\end{equation}
  In the following analysis, we will focus on the case where $T'=T+\v e_p \v e_q^\dag$; that is, changing an entry of $T$ from $0$ to $1$. The reverse case can be analysed similarly.
\section{Quantum and Quantum-Inspired Recommendation Algorithms}\label{sec3:algorithms}
  In this section we introduce the algorithmic models used for the quantum quantum-inspired classical recommendation algorithm~\cite{ewin_tang_ciq} in our analysis.
  For brevity, we refer to the latter as the quantum-inspired recommendation algorithm (or simply the quantum-inspired algorithm).
  
  Motivated by the low-rank reconstruction viewpoint described in Sec.~\ref{sec:recommen_prob}, both algorithms seek to compute $T_{\le k}$ efficiently and sample high-valued entry from its $i$-th row to recommend an item to user $i$. 
  
  More specifically, the quantum recommendation algorithm first computes and generates the quantum state $|{(T_{\le k})_i}\rangle$ then 
   \begin{algorithm}[H]
    \caption{Precise Singular Value Projection (adapted from \cite{q_recommend_sys})}
    \label{algo:q_svd_proj_precise}
    \begin{algorithmic}[1]
      \Require (Preference) matrix $T\in\mathbb R^{m\times n}$; Vector $\v x\in\mathbb{R}^n$ to be projected; Threshold parameter $\s\ge 0$.
      \Ensure $\ket{\v x}$ projected onto $\laspan \{\v v_\l:\sig_\l\ge\sig\}$, where $\ket{\v v_\l}$ are the left singular vectors of $T$ and $\s_\l$ are their corresponding singular values.

      \State Create $\ket\varphi\gets\ket{\v x}$ via the oracle provided by quantum-accessible classical database. Suppose $\ket{\v x}=\sum_{\l=1}^n \a_\l\ket{\v v_\l}$.
      \State Apply (precise) Quantum SVD on $\ket{\v x}$:
        \[\ket \varphi\gets\sum_\l\a_\l\ket{\v v_\l}\ket{\sig_\l}.\]
      \State Apply \textbf{Quantum If} condition on $\ket{\s_\l}$:
        \[\ket \varphi\gets
        \sum_{{\s}_\l\ge\s}\a_\l\ket{\v v_\l}\ket{\sig_\l}\ket{1}
        +\sum_{{\s}_\l<\s}\a_\l\ket{\v v_\l}\ket{\sig_\l}\ket{0}.
        \]
      \State Uncompute $\s_\l$:
        \[\ket\varphi\gets
        \sum_{{\s}_\l\ge\s}\a_\l\ket{\v v_\l}\ket{1}
        +\sum_{{\s}_\l<\s}\a_\l\ket{\v v_\l}\ket{0}.
        \]
      \State Measure on the second register. \textbf{If} the outcome is $1$, \textbf{Return} the first register.
      \State \textbf{Else}, goto step 1 and \textbf{repeat}.
    \end{algorithmic}
  \end{algorithm}
  \begin{algorithm}[H]
    \caption{Quantum Recommendation Algorithm (simplified, adapted from \cite{q_recommend_sys})}
    \label{algo:q_recommend_sys_model}
    \begin{algorithmic}[1]
      \Require Preference database matrix $T\in\mathbb R^{m\times n}$; User $i$ to recommend products.
      \Ensure A product $j$ recommended to user $i$.

      \State $k\gets$ some reasonable rank cutoff parameter that can seperate data from noise.
      \State $\ket\psi\gets\textsc{PreciseSingularValueProjection}(T,T_i,\sig_k)$.
      \State \textbf{Measure} $\ket\psi$ in the computational basis and get a product $j$.
    \end{algorithmic}
  \end{algorithm}
  \noindent measures it in the computational basis, so as to sample a product $j$ with high predicted rating $(T_{\le k})_{ij}$. For a given user $i$, the possibility that a product $j$ is recommended is
  \begin{equation}
      \Pr(j)=\frac{|(T_{\le k})_{ij}|^2}{|(T_{\le k})_i|^2}
  \end{equation} following the rule of quantum measurement.

  Under the hood, the quantum algorithm extends the quantum phase estimation algorithm \cite{nielsen_chuang_qcqi} to quantum singular value estimation (QSVE), and uses it to implement a singular-value filter that retains the components associated with the top singular values. Algo.~\ref{algo:q_svd_proj_precise}--\ref{algo:q_recommend_sys_model} provide a convenient pseudocode abstraction.

  \begin{algorithm}[t]
    \caption{Quantum-Inspired Recommendation Algorithm (adapted from \cite{ewin_tang_ciq})}
    \label{algo:ModFKV_}
    \begin{algorithmic}[1]
      \Require Matrix $T\in\mathbb R^{m\times n}$; Rank cutoff $k$; Error parameters $\eps, \kappa$.
      \Ensure (Description of) matrix $\td T_{\le k}$ that approximates $T_{\le k}$.

      \smallskip

      \State \textbf{(Calculate Parameters)}
      \State $K\gets |T|^2/\sig_k^2$, $\overline\eps\gets\kappa\eps^2$.
      \State Define size of downsampled matrix as $d\gets\Theta({K^4}/{\overline \eps^2})$.

      \smallskip
      \State \textbf{(Downsample)}
      \State Downsample rows of $T$ by $\l^2$-norm sampling, assemble chosen rows into $d\times n$ matrix $S$ with rescaling;
      \State Downsample cols of $S$ by $\l^2$-norm sampling, assemble chosen cols into $d\times d$ matrix $W$ with rescaling.

      \smallskip
      \State \textbf{(SVD and Filter)}
      \State Computer the top $k$ left singular vectors of $W$ $\v u_1, \v u_2,...,\v u_k$ that correspond to singular values $\sig_1,\sig_2,...,\sig_k$.
      \State Convert $\v u_\l \in \mathbb R^d$ into vectors $\v v_\l \in \mathbb R^n$ by
      \[
        \v v_\l = \frac{S^\dag\v u_\l}{|W^\dag \v u_\l|}.
      \]

      \State Output $\v v_1, \v v_2,...,\v v_k$ (the low rank approximation to $T$ is $\td T_{\le k}=T\sum_{\l=1}^k \v v_\l\v v_\l^\dag$).
    \end{algorithmic}
  \end{algorithm}

  Similarly, the quantum-inspired recommendation algorithm \cite{ewin_tang_ciq} generates an approximation to $T_{\le k}$ and samples its high-valued entries to provide recommendations. This algorithm is also based on SVD, but uses a probabilistic downsampling technique developed by Frieze et al. \cite{frieze2004fast} to achieve acceleration over traditional deterministic algorithms. 
  Inspired by the quantum counterparts, Tang's algorithm uses $\l^2$-norm sampling to mimic the distribution induced by quantum measurement, where an index is sampled with probability proportionate to the square norm of its associated vector component.
  Algo.~\ref{algo:ModFKV_} sketches the core procedure; further details are provided in Appendix~\ref{appendix:algo_detail} and the original literature.

  The error analysis of both algorithms are quite involved. For a clearer presentation and to isolate the privacy-relevant structure, we adopt a reasonably simplified model. Firstly, the error incurred by quantum phase estimation or matrix downsampling is ignored, as these errors do not have a big impact on our DP analysis; Next, we parametrize the singular-value filter by a rank cutoff $k$ instead of an explicit threshold $\sig$. 
  The reparametrization is after the fact that the cutoff is usually tuned by human operators such that the algorithm separates signal from noise; As aresult, different parametrizations would lead to equivalent behaviour after appropriate tuning. Formally, the simplified model of both algorithms can be defined as follows:

  \begin{definition}[Quantum and Quantum-Inspired Recommendation Algorithms]\label{def:simplified_algo_model} \hfill
    \begin{enumerate}
      \item The quantum recommendation algorithm $\A_{\rm RQ}^{k}$ takes in a preference database matrix $T$, a user index $i$, computes $\big|(T_{\le k})_i\big\>$ and measures the state in computational basis to get a product $j$ as its output.
      \item The quantum-inspired recommendation algorithm $\A_{\rm RC}^{k}$ takes in a preference database matrix $T$, a user index $i$, computes $T_{\le k}$ and samples a product $j$ from $i$-th row of $T_{\le k}$ by $\l^2$-norm sampling as its output.
    \end{enumerate}

    In both algorithms, $k$ is a rank cutoff parameter (tuned and chosen by human operators) such that the algorithm can separate data from noise. By the low-rank assumption, $k$ is in the order of $\rank(P)\sim\polylog\,(m,n)$, or $k=\O\,(\polylog\,(m,n))$. 
  \end{definition}

\section{Threat Model}\label{sec4:threat_model}
\begin{figure}[t]
  \centering
  \includegraphics[width=0.98\linewidth]{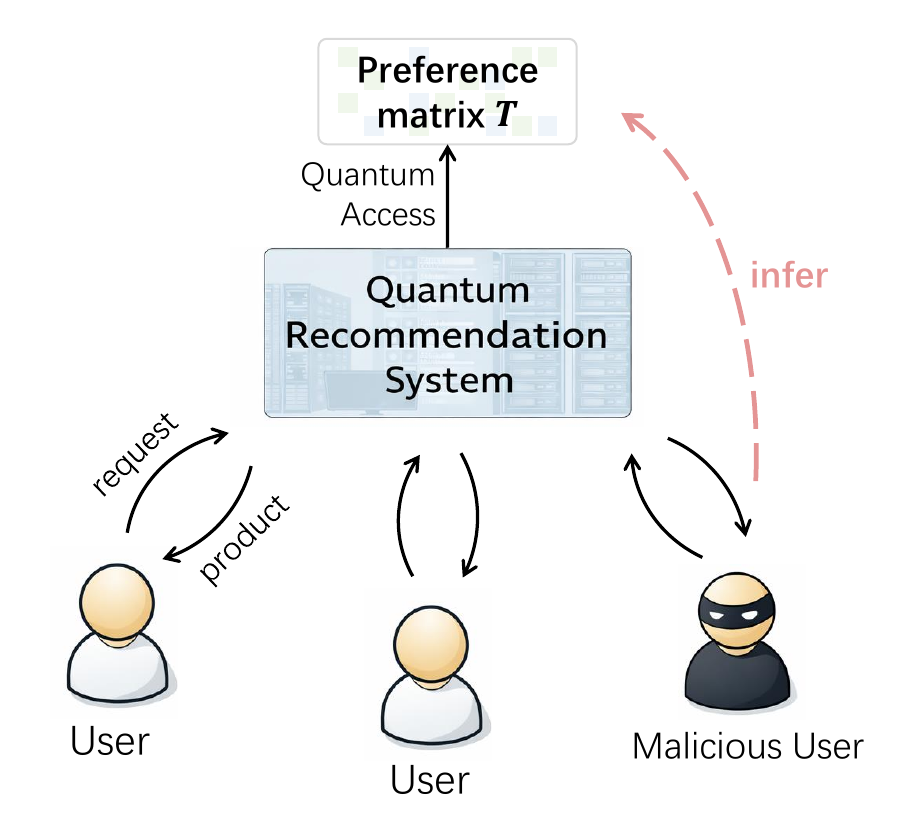}
  \caption{Illustration of our threat model. The server holds the private database $T$ and exposes a classical interface that maps a query user $i$ to a sampled recommendation $j$. The adversary observes only the outputs and may adaptively query and (optionally) contribute single rating records.}
  \label{fig:threat_model}
\end{figure}

  \paragraph{Centralized recommendation service.}
  The quantum (or quantum-inspired) recommendation algorithm provides an end-to-end recommendation service: given a user $i$, it outputs a recommended item $j$ in a randomized manner.
  Consequently, we model the quantum (or quantum-inspired) recommendation algorithm as a centralized recommendation server holding a private preference matrix $T$, as shown in Fig.~\ref{fig:threat_model}. Each invocation of the server specifies a user index $i$ and returns a single (random) recommended item $j$ sampled from the mechanism's output distribution.
  This interaction may be invoked repeatedly over time; however, we assume standard access-control measures such as authentication and rate limiting are in place to prevent unbounded querying.

  \paragraph{Adversary.}
  An adversary interacts with the server only through the above classical interface and observes the returned items.
  Its goal is to infer sensitive information about the underlying database, such as reconstructing parts of $T$ or learning particular entries of $T$.
We allow a strong adaptive adversary: it knows the algorithm and all public parameters and may choose queries adaptively based on past outputs.

  Importantly, the adversary's access is \emph{classical}: it observes only the sampled output item $j$ from each invocation.
  In particular, we do not assume the adversary can directly access the internal quantum state $\big|(T_{\le k})_i\big\rangle$ prior to measurement, nor any additional side information beyond the returned recommendation.

  \paragraph{Database updates and neighboring relation.}
  We assume the adversary can also \emph{contribute} records to the platform (e.g., by creating accounts and submitting ratings), thereby changing the database over time.
  To model privacy against such single-record changes, we adopt \emph{record-level} adjacency: neighboring databases $T\sim T'$ differ in exactly one entry $(p,q)$.
  Equivalently, for some $(p,q)$, one may write $T' = T \oplus \v{e}_p\v{e}_q^\dagger$ (entrywise addition modulo $2$) in the binary setting.
  
\section{Perturbation Methods for SVD}\label{sec5:method}
  Suppose a matrix $T$ corresponding to a recommendation dataset has SVD $T=\sum_\l \sig_\l \v u_\l \v v_\l^\dag$, then what is the SVD of its neighbour $T'=T\oplus \v e_p\v e_q^\dag=\sum_\l \sig_\l' \v u_\l' (\v v_\l')^\dag$, for a given pair of $(p,q)$? The answer to this question will be pivotal to our analysis over the differential privacy of the concerned algorithms and adversaries. Unfortunately, SVD is not numerically stable in terms of the singular vectors, and it is infeasible to express $\sig_\l',\v u_\l',\v v_\l'$ out of $\sig_\l,\v u_\l,\v v_\l$ in a closed-form manner in general. 

  In this section, we introduce our methods gathered from quantum mechanics \cite{zjy_qm_textbook, griffiths_qm_textbook}, random matrix theory \cite{feier_randmat, rand_ev_review, rand_mat_textbook} and matrix completion \cite{vandereycken2013low} to overcome this challenging problem. First we show that the added term $\delta T = \v e_p\v e_q^\dag$ is small enough to be viewed as a perturbation. Next, we improve the traditional perturbation method used in quantum mechanics to overcome the complicated numerical nature of SVD. Finally, we apply the perturbation method to neighbouring databases in the DP problem.

\subsection{Single Element Shift as a Perturbation}\label{sec:justify_pert}
  In this subsection, we show that the added term $\v e_p\v e_q^\dag$ is actually small enough to be viewed as a perturbation, opening ways to perturbation methods. 

  Suppose $T$ is a preference database matrix. Then it can be viewed as consisting of two components: the \emph{signal} component and the remained \emph{noise} component. The signal component captures the underlying patterns shared by users and items and is overall low-ranked; whereas the remaining component is dominated by noise, which is highly random and generally unstructured. Apparently, the strength of data should be much larger than that of noise so that they can be identified and recovered. Note that norm of SVD components is determined by $|\sum_\l\sig_\l\v u_\l\v v_\l^\dag|^2=\sum_\l\sig_\l^2$ for orthonormal basis $\{\v u_\l\}, \{\v v_\l\}$, so the strength of different components can be characterized by the magnitude of the corresponding singular values.

  Leveraging results from random matrix theory \cite{feier_randmat}, the following theorem characterizes the spectrum of random noise:

  \begin{theorem}[Marcenko-Pastur \cite{feier_randmat}, Singular Value Distribution of Noise]
    Suppose $N$ is an  $m\times n$ random matrix representing some noise. That is, each entry of $N$ is a random variable i.i.d. of mean $0$ and variance $1$, with $k$-th moment $\E[N_{ij}^k]<\infty$ not dependent on $n$. Then the singular value distribution of $N$ converges to the following distribution as $m\to\infty$ while keeping $n/m=:\a$ constant:
    \begin{align}
      p_\a(x)= \left\{\begin{array}{ll}
        \frac{\sqrt{(\sig_+^2-x^2)(x^2-\sig_-^2)}}{\pi x} & \sig_-\le x\le\sig_+ \\
        0 & \mathrm{Otherwise}
      \end{array}\right.
    \end{align}
    where $x$ corresponds to $\sig(N)/\sqrt{m}$ and $\sig_{\pm}:=1\pm\sqrt{\a}$.
  \end{theorem}

  From the theorem above, we see that the spectrum of random noise component with unit variance is confined to the interval $[\sqrt{m}\sig_-, \sqrt{m}\sig_+]=[\sqrt{m}(\sqrt{\a}-1),\sqrt{m}(\sqrt{\a}+1)]$. Consequently, the spectrum of random noise with intensity $I$ is lower bounded by $\sqrt{m}(\sqrt{\a}-1)I$. Assuming $m$ and $n$ are not too close and fix $I$, we have the lower bound $\sqrt{m}(\sqrt{\a}-1)I\sim\O(\sqrt{m})\gg 1$. 
  
  Furthermore, notice that the spectrum of the added term is $|\delta T|=|\v e_p\v e_q^\dag|=1$, we can establish the following hierarchy of strength of different components:
  \begin{multline}
    \{\text{strength of data}\}\gg\\
    \{\text{strength of noise}\}\gg\\
    \{\text{strength of } \delta T(=1)\}.\label{eq:hierachy_of_singval}
  \end{multline}
  This hierarchy reveals that the strength of the $\v e_p \v e_q^\dag$ term is even smaller than random noises, which is further much weaker than the strength of data. Therefore, it is reasonable to treat $\delta T = \v e_p \v e_q^\dag$ as a perturbation.

  \subsection{Low Rank Perturbation Ansatz for SVD}\label{sec:lora_pert_form}

  Perturbation methods are been widely used in quantum mechanics \cite{schrodinger1926quantisierung,zjy_qm_textbook,cohen_qm_textbook,griffiths_qm_textbook} to capture small changes on operators or matrices.
  However, the traditional form of perturbation \cite{griffiths_qm_textbook,zjy_qm_textbook} is unable to capture the ``sharp spike'' $\v e_p\v e_q^\dag$ and does not exhibit stable behaviour when applied to SVD. To tackle this problem, we propose a new perturbation ansatz as follows.

  We start with a standard first-order model for approximation. Suppose $T\to T'=T+\delta T$ where $\delta T=\lambda \v e_p\v e_q^\dag$ is a perturbation controlled by parameter $\lambda$, then we have:
\begin{equation}
T'_{\le k}=
T_{\le k} + \sum_{\ell\le k}\Big(
\delta\sigma_\ell\,\v{u}_\ell\v{v}_\ell^\dagger
+\sigma_\ell\,\delta\v{u}_\ell\,\v{v}_\ell^\dagger
+\sigma_\ell\,\v{u}_\ell\,\delta\v{v}_\ell^\dagger
\Big)+\O(\lambda^2),
\end{equation}
where $\delta\sigma_\ell,\delta\v{u}_\ell,\delta\v{v}_\ell$ capture the leading response to $\delta T$. Under the incoherence condition of $T$, it turns out that the first order correction for singular value is $\delta \sigma_\l=\v u_\l^\dag\delta T\v v_\l=\O(\frac{\lambda}{\sqrt{mn}})$, asymptotically smaller than the correction the for singular vectors, which are of orders $\O(\frac{\lambda}{\sqrt{m}})$ and $\O(\frac{\lambda}{\sqrt{n}})$, respectively. Consequently, we can focus on the corrections for the singular vectors rather than that for the singular values. Next, for a \emph{single-entry} update $\delta T=\lambda\v{e}_p\v{e}_q^\dagger$, the driving term is localized at row $p$ and column $q$.
This motivates a structured ansatz in which the dominant changes of singular vectors are also localized along $\v{e}_p$ and $\v{e}_q$.

Formally, we introduce the following perturbation ansatz, designed to capture the low-ranked ``sharp spike'' $\v e_p\v e_q^\dag$:
  \begin{equation}\label{eq:lora_pert_form}
T'_{\le k}
\approx
S'_{\le k}
:=
\sum_{\l\le k}\sig_\l\,
(\v{u}_\l+\a_\l \v{e}_p)\,
(\v{v}_\l+\b_\l \v{e}_q)^\dagger,
\end{equation}
  where $\a_\l,\b_\l\in\mathbb R$ are small perturbation coefficients. As shown in Theorem~\ref{thm:validity_of_pert_form}, these coefficients scale as $\O(\lambda/\sqrt{m}), \O(\lambda/\sqrt{n})$, respectively, asymptotically small in terms of the database size.
  
  To show the effectiveness of the proposed perturbation ansatz, we have the following theorem which bounds the residual error not captured by $S'_{\le k}$. Closer examinations behind the theorem reveal that the first order correction for singular vectors $\delta\v u_\l,\delta \v v_\l$ are indeed primarily focused on $\v e_p$ and $\v e_q$.

  \begin{theorem}\label{thm:validity_of_pert_form}
      Suppose $T$ is an incoherent matrix with SVD $T=\sum_\l\sig_\l\v u_\l\v v_\l^\dag$, and $T'=T+\delta T$ where $\delta T=\lambda \v e_p\v e_q^\dag$ is a perturbation controlled by parameter $\lambda$. Define the low rank perturbation ansatz as $S'_{\le k}:=\sum_{\le k}\sig_\l(\v u_\l+\a_\l\v e_p)(\v v_\l+\b_\l\v e_q)^\dag$. Then $S'_{\le k}$ captures the dominant correction of $\delta T$ for parameter choice $\a_\l=\lambda v_{\l q}/\sig_\l,\b_\l=\lambda u_{\l p}/\sig_\l$. More specifically, we have 
      \begin{equation}
      T'_{\le k}=S'_{\le k}+o(\lambda)=S'_{\le k}+O\left(\frac{\lambda k}{\sqrt{mn}}\right)+O(\lambda^2).
      \end{equation}
  \end{theorem}

  The proof is based on low-rank matrix projection and an analysis of the corresponding tangent space, and is therefore relatively technical. For the sake of brevity, we defer the full proof to Appendix~\ref{append:proof_1}.

\section{Differential Privacy of the Algorithms}\label{sec6:dp_res}
  In this section we present our main result on the differential privacy properties of the quantum and the quantum-inspired recommendation algorithms, and discuss their implications.
  
\subsection{Main Result}
  Based on the aforementioned data model and perturbation method, we can derive the following result for recommendation algorithms:

  \begin{theorem}[Main Result]\label{thm:main_thm}
For low-rank and incoherent input matrices, the quantum recommendation algorithm $\A_{\rm RQ}^k$ and the quantum-inspired recommendation algorithm $\A_{\rm RC}^k$ satisfy
\begin{enumerate}
  \item $\Big(\O\big(\sqrt{k/n}\big),\, \O\big(k^2/\min^2\{m,n\}\big)\Big)$-DP;
  \item in the typical regime $k=\polylog(m,n)$, equivalently
  \[\Big(\tilde{\O}(1/\sqrt{n}),\, \tilde{\O}(1/{\min}^2\{m,n\})\Big)\text{-DP},\]
\end{enumerate}
where $\tilde{\O}(\cdot)$ hides $\polylog(m,n)$ factors.
\end{theorem}

  Next we present the proof for our main result. For a clearer presentation, the technical details of order estimations in the proof are postponed to Appendix \ref{append:proof_order_est}.

  \paragraph{Proof} To prove our main result, it suffices to show that
  \begin{align}
    \Pr\big(\A_{\rm R}^{k}(T';i)=j\big)
    \le e^\veps \Pr\big(\A_{\rm R}^{k}(T;i)= j\big) + \delta
    \label{eq:line1}
  \end{align}
  holds for the desired values of $\veps, \delta$ for all users $i$ and all products $j$ on neighbouring databases $T$ and $T'$, where $\A_{\rm R}^k$ represents either $\A_{\rm RQ}^k$ or $\A_{\rm RC}^k$. Without loss of generality, suppose $T_{pq}=0$ and let $T'=T+\delta T=T+\v e_p\v e_q^\dag$. The other side where $T_{pq}=1\to 0$ can analysed similarly.

  Recall that both $\A_{\rm RQ}^{k}$ and $\A_{\rm RC}^{k}$ do $\l^2$-norm sampling from a row of the low-rank approximation $(T_{\le k})_i$ to provide recommendations to user $i$. So the probabilities expand to:
  \begin{gather}
   \Pr\big(\A_{\rm R}^{k}(T';i)=j\big) = \frac{\big|(T'_{\le k})_{ij}\big|^2}{\big|(T'_{\le k})_{i}\big|^2}, \\
   \Pr\big(\A_{\rm R}^{k}(T;i)=j\big) = \frac{\big|(T_{\le k})_{ij}\big|^2}{\big|(T_{\le k})_{i}\big|^2} . \label{eq:line12}
  \end{gather}

  For the sake of brevity, we will abbreviate the low-rank approximation $T_{\le k}$ and $T'_{\le k}$ as $\td T$ and $\td T'$ in the proof below, respectively.  Then we have the following relation:
  \begin{align}
    &\phantom{==}\Pr\big(\A_{\rm R}^{k}(T';i)=j\big)
    = \frac{|\td T'_{ij}|^2}{|\td T'_i|^2}\notag\\
    &= \Pr\big(\A_{\rm R}^{k}(T;i)=j\big) \cdot \frac{|\td T_i|^2}{|\td T'_i|^2} + \frac{|\td T'_{ij}|^2-|\td T_{ij}|^2}{|\td T'_i|^2} .
    \label{eq:dp_prim_ineq}
  \end{align}

  For the multiplicative bound, we start with the following relation:
  \begin{equation}
      |\td T'_i|^2-|\td T_i|^2 \le\big(|\td T'_i|+|\td T_i|\big)\cdot |\td T'_i-\td T_i|=|\td T'_i|\cdot\O\big(\sqrt{\sfrac kn}\big)
  \end{equation}
  where the last equality follows from the estimation $|\td T'_i-\td T_i|=\O\big(\sqrt{\sfrac kn}\big)$, which is based on the incoherence condition and the perturbation ansatz. Then we have the following bound for the multiplicative term $\frac{|\td T_i|^2}{|\td T'_i|^2}$:
  \begin{align}
    \ln \frac{|\td T_i|^2}{|\td T'_i|^2}
    \le \ln\frac{|\td T'_i|^2+|\td T'_i|\cdot\O\big(\sqrt{\sfrac kn}\big)}{|\td T'_i|^2}
    = \O\big(\sqrt{\sfrac kn}\big). 
  \end{align}
  
  Next we proceed to the additive bound. Again based on the incoherence condition, we can estimate that
  \begin{gather}
      \td T_{ij}=\O\left(\frac{k}{\sqrt{mn}}\right)=\O\left(\frac{k}{\min\{m,n\}}\right), \\
      \td T'_{ij}-\td T_{ij}=\O\left(\frac{k}{\min\{m,n\}}\right).
  \end{gather}
  then we have
  \begin{equation}
      \frac{|\td T'_{ij}|^2-|\td T_{ij}|^2}{|\td T'_i|^2}
      =\frac{1}{|\td T_i'|^2}\cdot\O\left(\frac{k^2}{\min^2\{m,n\}}\right) =\O\left(\frac{k^2}{\min^2\{m,n\}}\right),
  \end{equation}
  where the last equality follows from the fact that for any preference database matrix $T$ converted from a recommendation database, there is at least one non-zero element in each row. For example, in the famous Netflix and MovieLens \cite{dataset_movielens} dataset each user has  rated at least 20 movies, resulting in a lower bound of $|T_i|^2\ge 20$. Thus we take the conservative estimation $|\td T'_i|^2=\Omega(1)$, resulting in the last equality. 
  
  Plug the two bounds back to the inequality \eqref{eq:dp_prim_ineq}, and we would have proved that the quantum and dequantized recommendation algorithms are $\Big(\O\big(\sqrt{k/n}\big),\, \O\big(k^2/\min^2\{m,n\}\big)\Big)$-DP.
  
  For practical recommendation algorithms based on low rank approximation, $k$ is usually taken around $\rank(T)=\polylog(m,n)$. Then hiding any $\polylog(m,n)$ factors with $\td \O$ notations would result in the fact that both algorithms also satisfies $\Big(\tilde{\O}(1/\sqrt{n}),\linebreak \tilde{\O}(1/{\min}^2\{m,n\})\Big)$-DP.  \qed

\subsection{Discussion}
  Next we discuss some implications of our main result.

  \paragraph{Intrinsic randomness as a privacy source.} Based on Theorem \ref{thm:main_thm}, first we notice that the quantum and quantum-inspired recommendation algorithms require no additional noise to be differentially private. Conventional data curating mechanisms typically inject external noise into the data to obscure them before releasing to achieve differential privacy; However, the quantum (or quantum-inspired) recommendation algorithm is a curating mechanism \textit{on its own}: privacy protection is just a side effect of its intrinsic computation; no extra noise is needed.
  
  The reason behind this is a magical combination of the incoherence data condition and the quantum randomness arises from the quantum algorithm. Under incoherence condition, a single-entry update has only a delocalized and asymptotically small influence on each of the low-rank components, which translates to a \emph{small} shift of the output distribution.
  On the other hand, the quantum measurement in the quantum algorithm (or the $\l^2$-norm sampling process in the quantum-inspired algorithm) happens to provide the randomness that is necessary for any DP curating mechanisms.
  When both conditions are satisfied, the quantum (or quantum-inspired) recommendation algorithm becomes a differentially-private algorithm on its own, with no extra noise needed.
  We call such mechanisms \textit{passive} mechanisms, or \textit{noise-free} mechanisms.
  
  While traditional DP mechanisms struggle on privacy-utility tradeoff and guarantee user privacy at the cost of degraded utility, the quantum and quantum-inspired recommendation algorithms are noise-free, and provide privacy guarantees with no cost of utility. Previous research \cite{yang2017Privacy_JL2} has also shown that recommendation utility does not always have to be sacrificed when the privacy guarantee is properly aligned with the algorithm design. Here in the context of a quantum algorithm, the intrinsic curating ability is made possible by the probabilistic nature of quantum algorithms. We believe that more quantum or quantum-inspired algorithms with this property will emerge. 

  \paragraph{Privacy scaling with problem size.}
  Our bounds $\veps=\tilde{\mathcal{O}}(1/\sqrt{n})$ and $\delta=\tilde{\mathcal{O}}(1/\min^2\{m,n\})$ (for $k=\polylog(m,n)$) are asymptotically decreasing in the dataset dimensions.
  This suggests that, in large-scale recommendation systems where low-rank structure and incoherence condition hold, the per-query privacy loss of the recommendation interface becomes smaller as the number of items/users grows.
  Previous research \cite{mcsherry2009Differentially} has shown that utility loss of a differentially private recommendation system decreases with the dataset size increases under a fixed level of privacy guarantee (fixed DP parameter). This is because a fixed amount of error is gradually dominated by an increasing amount of data as the size of the datasets increases. Our work poses a nicely duality to their result: with fixed utility (no utility loss in our case), the privacy-preserving ability of a recommendation algorithm appears to improve as the size of datasets increase.

  \paragraph{Quantum vs.\ quantum-inspired.}
  Note that the quantum-inspired algorithm solves the same underlying low-rank reconstruction task and induces essentially the same sampling distribution as the quantum algorithm up to approximation error.
  As a result, the quantum-inspired algorithm enjoys the same privacy guarantees as its quantum counterpart. For the quantum-inspired algorithm, the required randomness arises from the $\l_2$-norm sampling process, which closely mimics the quantum measurement process.
  
\section{Experiments and Comparison}\label{sec7:exp}
In this section we empirically evaluate the concrete privacy parameters implied by our Theorem \ref{thm:main_thm} on real-world datasets, and compare the quantum (and quantum-inspired) recommendation algorithms with representative classical differentially private recommendation approaches.
Our goals are threefold:
(1) instantiate the asymptotic bounds in Theorem~\ref{thm:main_thm} by calibrating the hidden constants on standard recommendation benchmarks;
(2) validate the predicted scaling behaviour of $(\veps,\delta)$ as a function of dataset size;
and (3) perform a \emph{privacy-matched} comparison against classical noise-injection baselines, illustrating how much injected noise is required to reach the same privacy level under the same interface and neighbouring relation.

\subsection{Evaluation}\label{sec:exp_evaluate_eps}
We evaluate our results on the widely used MovieLens datasets~\cite{dataset_movielens}.
MovieLens provides several dataset variants spanning multiple orders of magnitude in the number of users, items, and ratings, making it a natural testbed for examining the asymptotic behaviour in Theorem~\ref{thm:main_thm}.

  \begin{table*}
  \centering
  \caption{Differential Privacy of Quantum Recommendation Algorithm on Datasets of Different Sizes}
  \label{tab:dp_param}
  \begin{tabular}{c|ccc|ccc|c}
    \toprule
    Dataset & Ratings & $m$ & $n$ & $k$ & $\veps$ & $\delta$ & \makecell[c]{$\veps^*$ for DP-SGD \cite{friedman2016Differential} \\ to beat baseline} \\ 
    \midrule
    MovieLens-100k & 100k & 943 & 1682 & 23 & 0.70 & 0.02 & 5.5 \\
    MovieLens-1m & 1m & 6040 & 3952 & 79 & 0.85 & 0.015 & 2 \\
    MovieLens-10m & 10m & 71567 & 10681 & 140 & 0.69 & 0.0025 & 1.5 \\
    MovieLens-25m & 25m & 162541 & 62423 & 197 & 0.32 & 0.00026 & (not reported) \\
    
  \bottomrule
\end{tabular}

\end{table*}

\begin{table}
\centering
\caption{Parameter Choice Summary}
\label{tab:parameter_choice_summary}
\begin{tabular}{c|c}
    \toprule
    Parameter & Value \\ 
    \midrule
    $\mu$ & 9 \\
    $\eps_{\rm approx}$ & 0.5 \\
    $k$ & $\min_{k'} \text{ s.t. } |T-T_{\le {k'}}|^2\le \eps_{\rm approx} |T|^2$ \\
  \bottomrule
\end{tabular}

\end{table}

\paragraph{What we measure.}
Our theory characterizes record-level DP for the \emph{single-item output interface} (Sec.~\ref{sec4:threat_model}): given a user index $i$, the mechanism outputs one recommended item $j$.
Accordingly, we report the per-query DP parameters $(\veps,\delta)$ implied by our bounds.
We emphasize that these are \emph{privacy} quantities; the purpose of this section is not to optimize recommendation accuracy, but to quantify privacy under realistic dataset statistics and parameter choices.
(We discuss utility/accuracy-related issues in Sec.~\ref{sec:exp_compare}.)

\paragraph{Parameter determination.}
To convert the asymptotic bounds into concrete numbers, we need to (1) fix an incoherence parameter $\mu$ and (2) settle the rank cutoff $k$. The determined parameters are summarized in Table \ref{tab:parameter_choice_summary}.

We follow the notation in Sec.~\ref{sec:recommen_prob}: if the singular vectors satisfy
\begin{equation}\label{eq:mu_def_exp}
  |U_{ij}|\le \sqrt{\mu/m},
  \qquad
  |V_{ij}|\le \sqrt{\mu/n},
\end{equation}
then tracing constants in the proof of Theorem~\ref{thm:main_thm} yields an instantiation of the leading term $\veps \approx \sqrt{\frac{\mu k}{n}}$.
We estimate $\mu$ empirically by computing the SVD on different MovieLens variants and examining the empirical distribution of singular vector coordinates.
As shown in Fig.~\ref{fig:uv_stat}, more than $98\%$ of the left and right singular vector coordinates fall within the $3/\sqrt{m}$ and $3/\sqrt{n}$ range respectively, across the datasets we tested.
Motivated by this, we adopt the parameter as $\mu=3^2=9$.

Next we select the rank cutoff $k$ via an approximation criterion aligned with common recommendation practice.
Specifically, we choose
\begin{equation}\label{eq:k_rule_exp}
  k := \min\Big\{k' : \ |T - T_{\le k'}|^2 \le \veps_{\rm approx}\,|T|^2 \Big\},
\end{equation}
where $\veps_{\rm approx}\in(0,1)$ controls the reconstruction fidelity.
In the quantum recommendation algorithm~\cite{q_recommend_sys}, the reconstruction fidelity is also related to the success probability of generating a high-quality recommendation; bigger $\veps_{\rm approx}$ corresponds to better reconstruction and better recommendations, at the cost of increasing $k$.
In practice, matrix-factorization-based recommenders often use $k$ in the tens for small datasets and rarely exceed a few hundred even for very large datasets~\cite{friedman2016Differential,koren2009matrix}.
To reflect this regime, we fix $\veps_{\rm approx}=0.5$, which yields $k$ values consistent with these best practices. The resulted $k$ values are shown in Table~\ref{tab:dp_param}.

Finally, to account for the omitted higher-order terms and modelling approximations (e.g., ignoring higher order errors in Theorem \ref{thm:validity_of_pert_form} and ignoring QSVE approximation error), we report a conservative instantiation of $(\veps,\delta)$ by multiplying the resulted values by an additional factor of $2$.
Table~\ref{tab:dp_param} summarizes the resulted DP parameters.

\paragraph{Results on MovieLens.}
Table~\ref{tab:dp_param} reports $(\veps,\delta)$ across MovieLens sizes.
Several trends are worth highlighting.

\emph{(1) Moderate per-query $\veps$ at scale.}
Across the evaluated datasets, $\veps$ stays below $1$ and often substantially below $1$.
In DP literatures, values such as $\veps \le \ln 2 = 0.69$ to $\veps \le 3$ are commonly considered ``reasonable'' privacy levels.
Our per-query $\veps$ values fall naturally within this range.

\emph{(2) Asymptotic improvement with dataset size.}
Consistent with Theorem~\ref{thm:main_thm}, $\delta$ decreases monotonically as $m,n$ grow, and $\veps$ shows an overall decreasing trend at large scale.
We notice that for MovieLens-1m and MovieLens-10m the decreasing trend of $\veps$ is not pronounced, and for MovieLens-1m there is even a slight increase
in $\veps$. This can be attributed to the fact that, for these datasets the
number of users grows rapidly, leading to a significant increase in $k$,
while the number of movies grows more slowly, resulting in a relatively large $\sqrt{k/n}$ factor. 
Once $n$ becomes sufficiently large (e.g., MovieLens-25m), $\veps$ drops significantly, resulting in stronger privacy guarantees.

\paragraph{Reference points from classical DP matrix factorization.}
For context, we also include in Table~\ref{tab:dp_param} the smallest privacy budgets reported by Friedman and Berkovsky~\cite{friedman2016Differential} for a DP matrix-factorization approach to match the accuracy of a non-private baseline (``$\epsilon^*$ for DG-SGD to beat baseline'').\footnote{We use their reported ``privacy budget required to beat baseline'' values as a coarse reference point. This is not an exact one-to-one comparison because the training objectives, evaluation metrics, and interfaces differ; we therefore also provide a privacy-matched noise-level comparison in Sec.~\ref{sec:exp_compare}.}
These values can be interpreted as the privacy budgets at which explicit-noise approaches begin to recover recommendation accuracy comparable to a non-private model.
By comparison, the quantum and quantum-inspired algorithms achieve much better privacy level in the same situation.
This also motivates the privacy-matched comparisons which we introduce below.

\subsection{Comparison of Algorithms}\label{sec:exp_compare}

\begin{table*}[t]
\centering
\caption{Comparison of Different Recommendation Algorithms on Their Privacy and Noise Setting}
\label{tab:algo_compare}
\begin{tabular}{cccc}
  \toprule
  Algorithm & Differential Privacy & Noise/Sampling & Noise \\
  \midrule
  DP-CF~\cite{mcsherry2009Differentially}
    & $\frac54\cdot\frac{(2B+3B^2)}{b}$-DP
    & \makecell[c]{Inject Laplace noise\\ to covariance statistics}
    & $\mathrm{Laplace}(b)^{n\times n}$ \\
  \midrule
  DP-CF~\cite{mcsherry2009Differentially}
    & $\left(2.5\ln(2/\delta)\cdot\frac{(1+2\sqrt{2}) B^2}{\sig}, \delta \right)$-DP
    & \makecell[c]{Inject Gaussian noise\\ to covariance statistics}
    & $\mathcal{N}(0, \sig^2)^{n\times n}$ \\
  \midrule
  Input noise~\cite{friedman2016Differential}
    & $\frac{1}{b}$-DP
    & \makecell[c]{Inject Laplace noise\\ to each rating $T_{ij}$}
    & $\mathrm{Laplace}(b)^{m\times n}$ \\
  \midrule
  DP-SGD~\cite{friedman2016Differential}
    & $\frac{\#\mathrm{iter}}{b}$-DP
    & \makecell[c]{Inject Laplace noise\\ to gradients}
    & $\mathrm{Laplace}(b)^{\#\mathrm{iter}}$ \\
  \midrule
  Quantum~\cite{q_recommend_sys}
    & $\Big(\O(\sqrt{k/n}),\, \O(k^2/\min^2\{m,n\})\Big)$-DP
    & Quantum measurement
    & / \\
  \midrule
  Quantum-inspired~\cite{ewin_tang_ciq}
    & $\Big(\O(\sqrt{k/n}),\, \O(k^2/\min^2\{m,n\})\Big)$-DP
    & $\l_2$-norm sampling
    & / \\
  \bottomrule
\end{tabular}
\end{table*}

Next we compare the quantum and quantum-inspired algorithms against representative classical DP recommendation approaches under comparable privacy budgets.

\paragraph{Baselines and settings.}
Prior work on DP recommendation broadly follows two directions: collaborative filtering (CF) and matrix factorization (MF).
CF methods exploit similarities between users and items to generate predictions; MF methods typically learn low-dimensional latent representations by solving the regularized optimization problem
\begin{equation}\label{eq:mat_factor_loss}
  \min_{\v p_i,\v q_j}
  \sum_{i,j} (T_{ij}-\v p_i^\dag\v q_j)^2
  +\lambda\Big(\sum_i|\v p_i|^2+\sum_j|\v q_j|^2\Big)
\end{equation}
where $\v p_i,\v q_j\in\mathbb R^k$ are latent user/item vectors. Under this model, a user's preference is then predicted as $\hat T_{ij}=\v p_i^\dag \v q_j$.
SVD-based reconstruction can be viewed as a special case of the matrix factorization paradigm.

We focus on two representative works in either approach.
McSherry and Mironov \cite{mcsherry2009Differentially} proposed a differentially private CF algorithm which injects Laplace or Gaussian noise into aggregated statistics of the dataset 
to achieve differential privacy. On the other hand, Friedman and Berkovsky \cite{friedman2016Differential} were the first to introduce differential privacy into MF–based recommendation methods. 
They primarily studied two DP schemes: (1) input noise injection, which adds Laplace noise to each individual rating in the preference matrix; and (2) gradient noise injection, 
which injects Laplace noise to the gradient evaluated at each iteration of optimization.
The privacy statements, sampling procedure and noise level of both works are summarized in Table \ref{tab:algo_compare}. The differentially private CF algorithm is marked as DP-CF, and the gradient noise injection algorithm is marked as DP-SGD in the table. For the DP-CF algorithms, $B$ is a parameter used to cap and clamp any normalized ratings to lower query sensitivity, whose value is taken to be $B=1$ in the original work. 

\paragraph{Privacy-matched noise level comparison.}

A key distinction between quantum and classical recommendation algorithms is that classical DP baselines enforce privacy by explicit noise injection, whereas the quantum and quantum-inspired recommendation algorithms do not.
To compare these paradigms on an equal footing, we ask the following question:

\begin{quote}
{\emph{How much injected noise is required for classical mechanisms to match the same $(\veps,\delta)$ level achieved by the quantum (or quantum-inspired) mechanism?}}
\end{quote}

To instantiate this comparison, we first estimate the $(\varepsilon, \delta)$ value for the quantum (or quantum-inspired) recommendation algorithm for different database sizes using methods similar to our evaluation in Sec.~\ref{sec:exp_evaluate_eps}, then compute the noise level required by other mechanisms to achieve the same level of privacy. As for parameter choices, we set $m=2n$ and select $k$ following common MF practice so that $k$ values match values appear in practical deployments~\cite{koren2009matrix}.
For the quantum mechanism we compute $(\veps,\delta)$ using the same calibration procedure as in the previous section.
For DP-SGD we use $\#\mathrm{iter}=5$ iterations, following the same parameter setting as in~\cite{friedman2016Differential}. To ensure a fair comparison between the classical and quantum recommendation algorithms, the dynamic range of rating $\Delta r$ of the classical algorithms is also set to $1$ (i.e., the ratings are normalized to lie within a unit range). For mechanisms with only an $\veps$ guarantee, we match $\veps$; for mechanisms with $(\veps,\delta)$, we match both parameters.

\begin{figure}[t]
  \centering
  \includegraphics[width=0.95\linewidth]{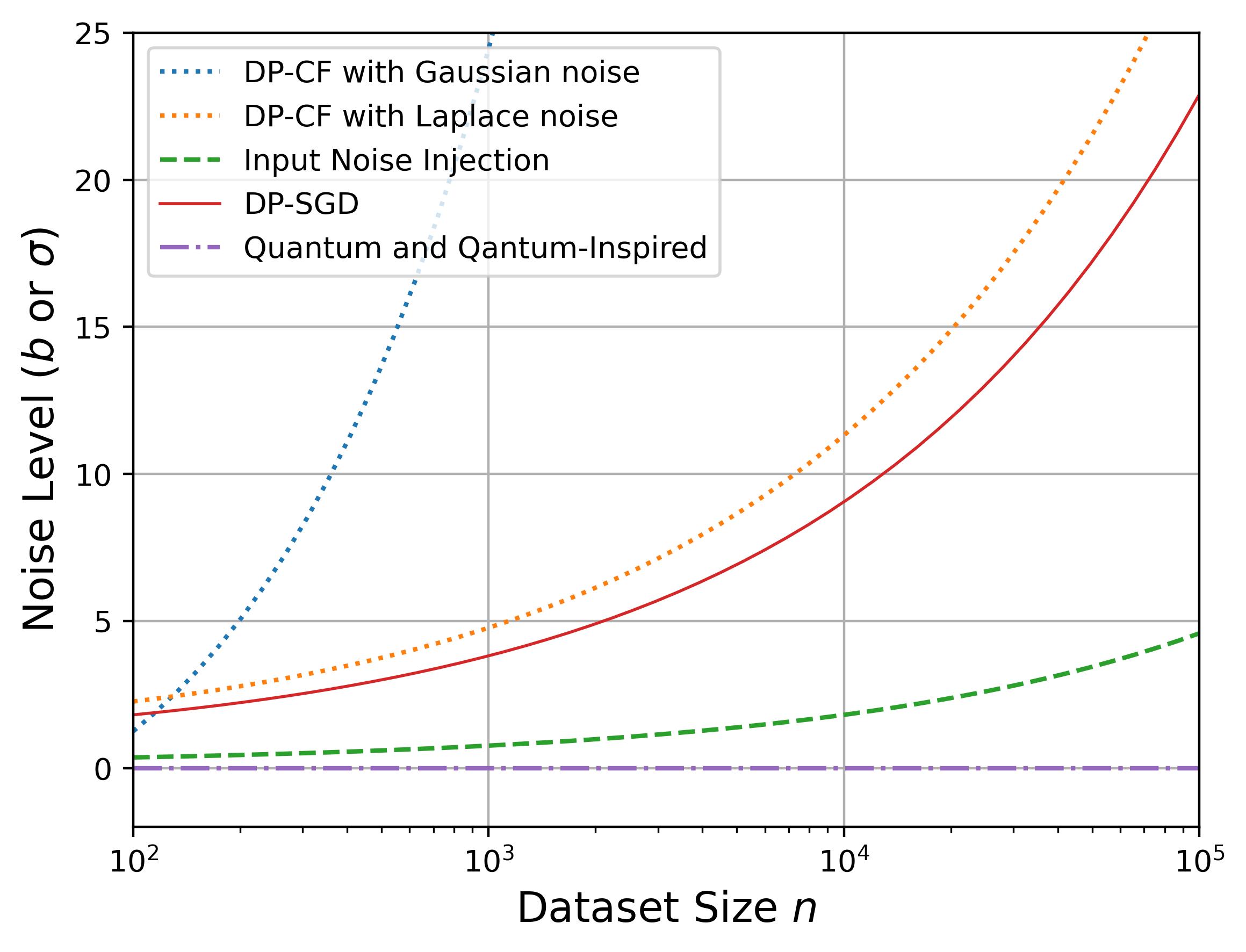}
  \caption{Noise level required for classical DP recommendation mechanisms to match the same privacy guarantee as the quantum mechanism, as a function of problem size.}
  \label{fig:equal_eps_noise_level_curve}
\end{figure}

The results are shown in Fig.~\ref{fig:equal_eps_noise_level_curve}, which reports the required noise scale as a function of problem size.
The trend is interesting: to match the same privacy level as the quantum (or quantum-inspired) mechanism, the magnitude of injected noise in classical baselines increases rapidly with $n$.
For instance, when $n=10^5$, the required entrywise Laplace noise scale approaches 5 for input noise injection, which is 5 times the full dynamic range of rating signal under our binary normalization! In such cases, we will have to drown data with noise to match the same level of privacy protection. The gap highlights the utility and scalability of the quantum recommendation algorithm in terms of privacy.

\subsection{Limitations}\label{sec:limitations}
Our experiments support the quantitative scaling predicted by Theorem~\ref{thm:main_thm} and highlight clear gaps against classical explicit-noise baselines under matched privacy levels.
We note a few practical considerations.

\paragraph{Modeling fidelity.}
We evaluate privacy under the simplified mechanism model in Def.~\ref{def:simplified_algo_model}, abstracting away implementation-level approximations (e.g., QSVE precision, downsampling error).
Fortunately, these effects do not change the asymptotic constants in our analysis and can be absorbed by the redundancy in privacy parameter estimation, or alternatively addressed through a more refined analysis.

\paragraph{Data conditions.}
Our guarantees leverage standard low-rank and incoherence conditions that are empirically plausible for many recommender datasets, but may degrade in regimes with high coherence (e.g., highly concentrated singular vectors, extreme user/item, or adversarially structured data), where single-entry updates can have larger influence.

\paragraph{Repeated interaction.}
The stated $(\veps,\delta)$ bounds are per invocation of the recommendation interface.
In practice, repeated queries incur composed privacy loss via standard DP composition, so the effective guarantee depends on the query budget.

\paragraph{Comparison scope.}
Our privacy-matched comparison emphasizes how much additional noise classical baselines must inject to reach the same privacy level.
A full utility benchmark across mechanisms under identical training and setting is complementary and left for future work.

\section{Related Works}\label{sec:related_works}
In this section we review the most relevant lines of research and position our research within them.

\paragraph{Differentially-private recommendation systems.} After the well-known Netflix Prize dataset was shown to be vulnerable to de-anonymization attacks \cite{narayanan2006deanonymize}, privacy concerns in recommendation systems gained widespread attention, motivating a line of research on differentially private recommendation algorithms.
McSherry and Mironov~\cite{mcsherry2009Differentially} propose DP mechanisms for collaborative filtering (CF) methods by adding calibrated Laplace/Gaussian noise to aggregated quantities such as item--item similarity or covariance matrices.
Subsequent work studies DP variants for neighbourhood-based recommendation, clustering-based recommenders, and matrix-completion-style interfaces under different adjacency notions and output interfaces~\cite{yang2017Privacy_JL2,berlioz2015Applying,shin2018privacy}.

Another influential direction is differentially private matrix factorization (MF), where recommendations are produced from low-dimensional user/item embeddings learned by minimizing an empirical loss.
Friedman and Berkovsky~\cite{friedman2016Differential} systematically investigate MF under DP and compare multiple noise injection strategies, including perturbing inputs and perturbing optimization dynamics (adding noise to gradients).
These approaches demonstrate the dominant classical paradigm: privacy guarantees are enforced by adding external noise, yielding a privacy--utility tradeoff controlled by noise magnitude.

Our work differs in both \emph{object} and \emph{mechanism}.
We analyze the quantum recommendation algorithm~\cite{q_recommend_sys} and its quantum-inspired counterpart~\cite{ewin_tang_ciq} under the standard record-level neighboring relation.
Most importantly, our DP guarantees arise from the \emph{intrinsic} sampling/measurement randomness already present in these algorithms, rather than from adding additional DP noise to the data or learned model.
This leads to an explicit $(\veps,\delta)$ characterization as a function of $m,n$ and $k$.

\paragraph{Quantum DP and privacy in quantum learning.}
The relationship between DP and quantum information has been studied from several angles.
Zhou and Ying~\cite{qdp_ying} initiate a formal treatment of differential privacy for quantum data.
Aaronson and collaborators connect DP results to gentle quantum measurements~\cite{aaronsonGentleMeasurementQuantum2019}, relating privacy guarantees on database to limits on information extraction on quantum states.
More recent work studies variants such as local quantum differential privacy and other privacy formalisms tailored to quantum channels and quantum data~\cite{nuradha2024quantum,guan2025optimal,hirche2023quantum}.

Most prior quantum-DP work focuses on privacy notions for general quantum datasets (states/channels).
In contrast, our research focus on an end-to-end quantum machine learning algorithm operating on classical datasets: the quantum state $|(T_{\le k})_i\rangle$ is an internal artifact, and the adversary observes only classical outputs through the recommendation interface.

\paragraph{Matrix completion and perturbation tools.}
Matrix completion and low-rank approximation are the mathematical foundations for modern matrix-factorization-based recommendation systems.
Cand\`es and Tao~\cite{candes2010power} formalize conditions (including incoherence) enabling recovery of low-rank structure from partial observations, and Tao~\cite{tao2010matrixcompletion} emphasizes incoherence/delocalization as necessary to avoid degenerate instances.
Incoherence also appears in DP analyses of matrix computations, where structural regularity can reduce sensitivity and improve privacy bounds~\cite{hardt2013beyond}. Vandereycken~\cite{vandereycken2013low} introduces Riemannian optimization and matrix manifolds into matrix completion, which plays an important role in our proof to Theorem \ref{thm:validity_of_pert_form}.

Classical perturbation results (e.g., Stewart/Wedin-type bounds) provide guarantees for singular values and subspaces under perturbations~\cite{stewart1998perturbation}, but directly tracking singular vectors can be unstable and complicated, especially when gaps are small. Instead, we follow the alternate approach involving matrix manifold and tangent space initiated by Vandereycken~\cite{vandereycken2013low}.
Our technical contribution in Sec.~\ref{sec5:method} is a spike-aware perturbation framework tailored to the DP neighbouring relation on incoherent matrices.
The resulted bounds control how the recommendation sampling distribution changes, enabling our end-to-end DP guarantees and the privacy-matched comparisons in Sec.~\ref{sec7:exp}.

\section{Conclusion and Future Work}\label{sec:conclusion}
We analyzed the differential privacy of the quantum recommendation algorithm $\mathcal{A}_{\rm RQ}^k$~\cite{q_recommend_sys} and the quantum-inspired recommendation algorithm $\mathcal{A}_{\rm RC}^k$~\cite{ewin_tang_ciq}.
Under record-level adjacency and the threat model articulated in Sec.~\ref{sec4:threat_model}, we show that both algorithms satisfy \linebreak
$\big(\td\O(1/\sqrt n),\ \td\O(1/\min^2\{m,n\})\big)$-DP
without injecting additional noise.
The guarantee follows from two key facts: (1) incoherence makes the overlap between a single-entry ``spike'' update and the leading singular directions asymptotically small, and (2) the algorithms' native sampling steps (quantum measurement / $\l_2$-sampling) provide exactly the randomness required by DP.
Technically, we develop a spike-aware perturbation method for truncated SVD that translates a one-record database change into explicit control of the induced recommendation distribution, yielding end-to-end DP guarantees for both algorithms.
Finally, we instantiate the resulted $(\veps,\delta)$ bounds on MovieLens datasets and observe scaling behaviour consistent with our theory.

\paragraph{Future work.}
We plan to (1) extend the privacy analysis to realistic implementations by accounting for QSVE/downsampling and finite-precision errors, (2) study user-level adjacency beyond single-entry perturbations, and (3) study the privacy of other quantum and quantum-inspired machine learning algorithms.

\section*{Acknowledgments}




The authors are grateful to Ke'an Chen, who helped us rule out the infeasible path of seeking an explicit solution to the SVD perturbation problem. We would like to thank Qi Zhou and Chengye Li for  providing us with some key references on random matrix theory. We also thank Chun'an Shi for his insightful discussions.

\appendix
\section*{Ethical Considerations}

\paragraph{Stakeholder Identification and Impacts.}
Our research involves three primary stakeholder groups as follows. (1) \emph{Platform operators / service providers (deployers of recommendation systems):} Our research may encourage platforms to deploy the quantum or quantum-inspired recommendation algorithm. The platforms and commercial companies may benefit from the noise-free recommendation algorithms, as they provide privacy guarantee without compromising utility, if the data condition is met. Conversely, they may be harmed by legal/regulatory or reputational risks if they over-claim privacy guarantees or ignore deployment constraints (e.g., repeated queries, non-incoherent data). (2) \emph{End users of recommendation systems:} They may benefit from stronger formal privacy guarantees on recommendation outputs. They may also be harmed if our results are misunderstood as “privacy for free” in settings where assumptions do not hold (e.g., high-coherence regimes) or where repeated queries are not properly accounted for. (3) \emph{Researchers and practitioners:} Our work may influence future research directions by motivating further study of privacy in quantum/quantum-inspired learning algorithms and by clarifying when intrinsic randomness (measurement or $\ell_2$-sampling) can act as the randomness required by differential privacy.

\paragraph{Potential Harms and Mitigation.}
We acknowledge that our research may create negative impacts if \emph{misinterpretation or misuse} happens to our theoretical guarantees, leading to exaggerated claims of “noise-free DP” or “quantum advantage” in deployments where prerequisites do not hold. In particular, privacy bounds rely on structural assumptions (low-rank and incoherence) and are stated for a specific interface and neighbouring relation (record-level adjacency and a single-item output per invocation). If a platform ignores these conditions (e.g., deploys on highly coherent data without checks) or ignores privacy loss under repeated interaction (DP composition), end users may receive weaker privacy protection than advertised. To mitigate this, we (1) explicitly and repeatedly state the low-rank and incoherence assumptions as prerequisites for our results, and (2) emphasize that the stated $(\varepsilon,\delta)$ guarantees are \emph{per invocation} and must be composed over multiple queries. We further encourage deployers to validate coherence/low-rank conditions on their own data before relying on the guarantees, and to implement standard access controls such as authentication, rate limiting, and auditing.

\paragraph{Respect for law and public interest.}
Our work is based on offline analysis and experiments on public datasets and does not involve interacting with live systems or bypassing access controls. We highlight deployment guidance (rate limiting, authentication, composition) that aligns with public interest in privacy-preserving services.

\paragraph{Human subjects.}
This work does not involve interventions with human participants, deception, or experiments on live services without consent. Our evaluation uses offline computation on existing public datasets; we only report aggregate results and do not release any user-level derived artifacts that would increase re-identification risk beyond what is already available in the original datasets.

\paragraph{Decision to proceed and to publish.}
We proceeded because the research procedures are low-risk (theoretical analysis and offline experiments) and the primary expected impact is positive: quantum and quantum-inspired recommendation algorithms may provide formal privacy guarantees without additional noise injection under clearly stated structural conditions. While assumptions on data are required, we believe that making these assumptions explicit, together with deployment suggestions that address foreseeable misuse (e.g., data-condition checks), preserves the public-interest benefits of open scientific communication.

\section*{Open Science}
We mainly use Python 3.10.9 with Numpy and Scipy to run experiments and Matplotlib to draw the results. The experiments are 
run on a PC with Intel Core i7-8750H CPU
and 16GB memory and the code should be compatible with
any recent versions.

The MovieLens datasets can be downloaded from \url{https://files.grouplens.org/datasets/movielens/}, and our code is available on via the anonymous link \url{https://anonymous.4open.science/r/paper-DP-for-qRS_artifacts-82B0}.
The components of the anonymous repository are as follows:
\begin{itemize}
    \item \verb|README.md|: The readme file.
    \item \verb|incoherence_demo.ipynb|: Jupyter notebook that checks the incoherence of MovieLens datasets and plots Fig.~\ref{fig:uv_stat}.
    \item \verb|exp_evaluation.ipynb|: Jupyter notebook that conducts the experiments and evaluation in Sec.~\ref{sec:exp_evaluate_eps} and computes Table~\ref{tab:dp_param}.
    \item \verb|privacy_matched_noise_level_comparison.ipynb|: Jupyter notebook that plots Fig.~\ref{fig:equal_eps_noise_level_curve}.
    \item \verb|data_loader.py|: Data loader that loads certain MovieLens dataset into matrix form. The datasets need to be be in the same folder as the loader script.
\end{itemize}
\cleardoublepage

\bibliographystyle{plainurl}
\bibliography{libpaper}

\section{Deferred Proof}\label{append:proof}

\subsection{Proof of Theorem \ref{thm:validity_of_pert_form}}\label{append:proof_1}
    Define the low rank matrix space as
    \begin{equation}
        \mathcal M_k:=\{M\in \mathbb R^{m\times n}\, |\, \rank(M)\le k\},
    \end{equation}
    which is known to be a smooth and differentiable manifold \cite{vandereycken2013low}.
    We further define the projection function onto it
    \begin{equation}
        \Pi_{\mathcal M_k}(X):=\argmin_{\td X\in\mathcal M_k}|X-\td X|,\label{eq:opt_prob_thm_proof}
    \end{equation}
    then the Eckart-Young theorem tells us that $\Pi_{\mathcal M_k}(T)=T_{\le k}, \Pi_{\mathcal M_k}(T+\delta T)=T'_{\le k}$. 
    
    Define the tangent space of $\mathcal M_k$ at $T_{\le k}$ as $\mathrm{T} \mathcal M_k(T_{\le k})$, then moving along any direction in $\mathrm{T} \mathcal M_k(T_{\le k})$ does not impact the residual error up to first order; otherwise $T_{\le k}$ would not be the optimal solution to \eqref{eq:opt_prob_thm_proof}. Specifically, the tangent space of the low rank matrix manifold can be expressed as (\cite{vandereycken2013low}, Prop. 2.1):
    \begin{equation}
        \mathrm{T}\mathcal M_k({T\lek}) = \Big\{\dd T_{\le k}\,:\, \dd T_{\le k}=\sum_{\l\le k}\v u_\l\dd \v v_\l^\dag+\dd \v u_\l\v v_\l^\dag\Big\}.
    \end{equation}
    Intuitively, matrices in the tangent space must be linearly dependent to either $\{\v u_\l\}$ or $\{\v v_\l\}$ to blend into the existing rank components; the matrices cannot be linearly independent to both sides as that would introduce an extra rank. Equivalently, the tangent space can be expressed as
    \begin{align}
        \mathrm{T}\mathcal M_k({T\lek}) &= \Big\{\dd M\,:\, \dd M=(I_m-\Pi_k^U)M(I_n-\Pi_k^V),\, M\in\mathbb R^{m\times n}\Big\}
    \end{align}
    where the projectors $\Pi_k^U,\Pi_k^V$ are defined as $\Pi_k^U:=\sum_{\l\le k}\v u_\l\v u_\l^\dag$ and $\Pi_k^V:=\sum_{\l\le k}\v v_\l\v v_\l^\dag$.
    
    When $T$ shifts to $T'=T+\delta T$, the correction of $T_{\le k}$ primarily focuses on the tangent space of $\mathcal M_k$ at $T_{\le k}$ for small $\delta T$. That is to say, the contribution of $\delta T$ onto $T_{\le k}$ can be expressed as
    \begin{equation}
        T_{\le k}'=T_{\le k}+\Pi_{\mathrm{T}\mathcal M_k({T\lek})}(\delta T)+\O(\lambda^2)
    \end{equation}
    where $\Pi_{\mathrm{T}\mathcal M_k({T\lek})}(\delta T)=\Pi_k^U \delta T+\delta T\Pi_k^V-\Pi_k^U\delta T\Pi_k^V$ is the projection onto the tangent space. Then we further have the following result for the terms in the projection $\Pi_{\mathrm{T}\mathcal M_k({T\lek})}(\delta T)=\Pi_k^U \delta T+\delta T\Pi_k^V-\Pi_k^U\delta T\Pi_k^V$:
    \begin{gather*}
        \Pi_k^U\delta T 
          = 
        \lambda\Big(\sum_{\l\le k} \v u_\l \v u_\l^\dag\v e_p \Big)\v e_q^\dag
          =
        \sum_{\l\le k}\underbrace{(\lambda u_{\l p})}_{\bar\b_\l}\v u_\l\v e_q^\dag = \O\Big(\frac{\lambda k}{\sqrt{m}}\Big),\\
        \delta T \Pi_k^V
          = 
        \lambda\v e_p\Big(\sum_{\l\le k} \v e_q^\dag \v v_\l \v v_\l^\dag \Big)
          =
        \sum_{\l\le k}\underbrace{(\lambda v_{\l q})}_{\bar\a_\l}\v e_p\v v_\l^\dag=\O\Big(\frac{\lambda k}{\sqrt{n}}\Big),\\
        \Pi_k^U\delta T \Pi_k^V
          = \lambda\Big(\sum_{\l\le k} u_{\l p}\v u_\l\Big)\Big(\sum_{\l'\le k} v_{\l' p}\v v_{\l '}\Big) = \O\Big(\frac{\lambda k}{\sqrt{mn}}\Big).
    \end{gather*}
    
    Then by collecting the first and second term and adding the cross term $\sum_{\l\le k}\sig_\l\a_\l\b_\l\v e_p\v e_q=\O\big(\frac{\lambda^2 k}{\sqrt{mn}}\big)$, we have got the low rank ansatz $S'_{\le k}$. More specifically, we take $\bar\a_\l=\lambda u_{\l p},\bar\b_\l=\lambda v_{\l q}$ where $\bar \a_\l,\bar \b_\l$ are defined as $\bar \a_\l=\sig_\l\a_\l,\bar\b_\l=\sig_\l\b_\l$. At the meantime, the residual error is
    \begin{align}
        T'_{\le k}-S'_{\le k}&= \O(\lambda^2)+\Pi_k^U\delta T\Pi_k^V-\sum_{\l\le k}\sig_\l\a_\l\b_\l\v e_p\v e_q \\
        &= \O(\lambda^2)+\O\Big(\frac{\lambda k}{\sqrt{mn}}\Big)-\O\Big(\frac{\lambda^2 k}{\sqrt{mn}}\Big) \\
        &= \O(\lambda^2)+\O\Big(\frac{\lambda k}{\sqrt{mn}}\Big).
    \end{align}
  \qed

  \subsection{Order Estimations in the Proof of Main Result}\label{append:proof_order_est}
  \paragraph{Estimate $|\td T'_i-\td T_i|$.}  Define $\Delta_{\le k}:=T'_{\le k}-T_{\le k}$, then we begin by the estimation of $\Delta_{\le k}=T'\lek-T\lek\approx S'\lek-T\lek$. By low rank perturbation ansatz, we have $T'_{\le k}-T_{\le k}\approx S'_{\le k}-T_{\le k}=\sum_{\l \le k}\sig_\l(\a_\l\v e_p\v v_\l^\dag+\b_\l\v u_\l\v e_q^\dag+\a_\l\b_\l\v e_p\v e_q^\dag)\approx \sum_{\l\le k}\bar\a_\l\v e_p\v v_\l^\dag+\bar\b_\l\v u_\l\v e_q^\dag$ after ignoring the higher-order terms. Then we have:
  \begin{align*}
    |(\Delta_{\le k})_i|^2&\cong\left|\v e_i^\dag\sum_{\l\le k}\bar\a_\l\v e_p\v v_\l^\dag+\bar\b_\l\v u_\l\v e_q^\dag\right|^2\\
    &=\left|\sum_{\l\le k}\delta_{ip}\bar\a_\l\v v_\l+\bar\b_\l u_{\l i}\v e_q^\dag\right|^2\\
    &\le\left|\sum_{\l\le k}\bar\a_\l\v v_\l\right|^2+\left|\sum_{\l\le k}\bar\b_\l u_{\l i}\v e_q^\dag\right|^2\\
    &=\sum_{\l\le k}\bar\a_\l^2+\sum_{\l\le k}(\bar\b_\l u_{\l i})^2\\
    &=\sum_{\l\le k}\vlj^2+u_{\l p}^2 u_{\l i}^2=\O(\sfrac{k}{n}). 
  \end{align*}
  Then taking the square root on both sides gives $|\td T'_i-\td T_i|=\O(\sqrt {k/n})$.

  \paragraph{Estimate $\td T_{ij}$.} By the SVD of $T$ we have $(T_{\le k})_{ij}=\v e_i\big(\sum_{\l\le k}\sig_\l\v u_\l\v v_\l^\dag\big)\v e_j^\dag=\sum_{\l\le k}\sig_\l u_{\l i} v_{\l j}=\O\big(\frac{k}{\sqrt{mn}}\big)$, where the last inequality follows from the incoherence condition of $T$.

  \paragraph{Estimate $\td T'_{ij}-\td T_{ij}$.} Again define $\Delta_{\le k}:=T'_{\le k}-T_{\le k}\approx S'_{\le k}-T_{\le k}=\sum_{\l \le k}\sig_\l(\a_\l\v e_p\v v_\l^\dag+\b_\l\v u_\l\v e_q^\dag+\a_\l\b_\l\v e_p\v e_q^\dag)$, then we can derive the following result for matrix entries of $\Delta_{\le k}$ after ignoring the higher-order terms caused by replacing $T'_{\le k}$ with the perturbation ansatz
  \begin{multline}
      (\Delta_{\lek})_{ij}\cong
      \sum_{\l \le k}\Big\{
      \delta_{ip}v_{\l q}v_{\l j}+
      \delta_{jq}u_{\l p}u_{\l i}+
      \delta_{ip}\delta_{jq}u_{\l p}v_{\l q}/\sig_\l
      \Big\}
  \end{multline}
  where $\delta_{ij}$ is the Kronecker $\delta$ notation:
  \begin{equation}
      \delta_{ij}=\left\{\begin{array}{ll}
          1 & i=j \\
          0 & i\neq j
      \end{array}\right. .
  \end{equation}
  Then by the incoherence of matrix $T$ and the estimation of rank cutoff parameter $\sig_k\ge 1$ we have
  \begin{equation}
      (\Delta_{\le k})_{ij}= k\cdot \O\left(\max\left\{\frac1n,\frac1m,\frac{1}{\sqrt{mn}}\right\}\right)=\O\left(\frac{k}{\min\{m,n\}}\right).
  \end{equation}
  \newpage

\section{Details of Quantum and Quantum-Inspired Classical Recommendation Algorithms}\label{appendix:algo_detail}

  In this section we show the original quantum and quantum-inspired recommendation algorithm in detail, in case readers may need to refer to them.

  First we review the quantum recommendation algorithm and the singular value projection subroutine required by it.

  \begin{algorithm}[H]
  \caption{Singular Value Projection}
  \label{algo:q_svd_proj}
  \begin{algorithmic}[1]
      \Require (Preference database) Matrix $T\in\mathbb R^{m\times n}$; Vector $\v x\in\mathbb{R}^n$ to be projected; Threshold parameter $\s\ge 0$.
      \Ensure $\ket{\v x}$ projected onto $\laspan \{\v v_\l:\sig_\l\ge\sig\}$ (with some error), where $\ket{v_\l}$ are the left singular vectors of $T$ and $\s_\l$ are their corresponding singular values. In other words, return $\ket{\psi}=\sum_{\td \s_\l\ge\s}\a_\l\ket{\v v_\l}\cong \sum_{\s_\l\ge\s}\a_\l\ket{\v v_\l}$.

    \bigskip

    \State Create $\ket\varphi\gets\ket{\v x}$ via the oracle provided by quantum-accessible classical database. Suppose $\ket x=\sum_{i=1}^n \a_\l\ket{v_\l}$.
    \State Apply Quantum SVD on $\ket \varphi$:
      \[\ket {\varphi}\gets\sum_\l\a_\l\ket{\v v_\l}\ket{\td\sig_\l}\]
    \State Apply \textbf{Quantum If} condition on $\ket{\td\s_\l}$:
      \[\ket\varphi\gets
      \sum_{\td{\s}_\l\ge\s}\a_\l\ket{\v v_\l}\ket{\td\sig_\l}\ket{0}
      +\sum_{\td{\s}_\l<\s}\a_\l\ket{\v v_\l}\ket{\td\sig_\l}\ket{1}
      \]
    \State Uncompute $\td \s_\l$:
      \[\ket\varphi\gets
      \sum_{\td{\s}_\l\ge\s}\a_\l\ket{\v v_\l}\ket{0}
      +\sum_{\td{\s}_\l<\s}\a_\l\ket{\v v_\l}\ket{1}
      \]
    \State Measure on the last register. \textbf{If} the outcome is $0$, output the first register and quit.
    \State \textbf{Else}, goto step 1 and \textbf{repeat}.
  \end{algorithmic}
  \end{algorithm}

  \begin{algorithm}[H]
    \caption{Quantum Recommendation Algorithm \cite{q_recommend_sys}}
    \label{algo:q_recommend_sys}
    \begin{algorithmic}[1]
      \Require (subsampled) Preference database matrix $T\in\mathbb R^{m\times n}$; User $i$ to recommend products.
      \Ensure A product $j$ recommended to user $i$.

      \State $\sig\gets$ some reasonable threshold that can seperate data from noise.
      \State $\ket\psi\gets\textsc{SingularValueProjection}(T,T_i,\sig)$.
      \State \textbf{Measure} $\ket\psi$ in the computational basis and get a product $j$.
    \end{algorithmic}
  \end{algorithm}

  The expected runtime of the quantum recommendation algorithm is $\O\big(\polylog(mn)\cdot\frac{|T|}{\sig_k}\big)$, where $\sig_k$ is the singular value threshold.

  Based on identity $|T|^2=\sum_\l \sig_\l^2$ where $\sig_l$ are all singular values of $T$, it can be estimated that $\sig$ should be in order of $\frac{|T|}{\sqrt k}$, where $k$ is the rank cutoff parameter mentioned in Def. \ref{def:simplified_algo_model} and is usually taken near $\rank(T)$. Then the expected time complexity of the quantum recommendation algorithm can be simply expressed as $\O(\sqrt k\cdot \polylog(mn))$.

  Next we show the quantum-inspired classical algorithm proposed by Tang \cite{ewin_tang_ciq}.

  \begin{algorithm}[H]
    \caption{Quantum-Inspired Classical Recommendation Algorithm (adapted from \cite{ewin_tang_ciq})}
    \label{algo:ModFKV}
    \begin{algorithmic}[1]
      \Require Matrix $T\in\mathbb R^{m\times n}$; Threshold parameter $\sig$; Error parameters $\eps, \kappa$.
      \Ensure (Description of) matrix $\td T_{\le k}$ that approximates $T_{\le k}$.

      \smallskip

      \State \textbf{(Calculate Parameters)}
      \State $K\gets |T|^2/\sig^2$, $\overline\eps\gets\kappa\eps^2$.
      \State Define size of downsampled matrix as $q\gets\Theta({K^4}/{\overline \eps^2})$.

      \smallskip
      \State \textbf{(Downsample)}
      \State Downsample rows of $T$ by $\l^2$-norm importance sampling, assemble chosen rows into $q\times n$ matrix $S$ with rescaling;
      \State Downsample cols of $S$ by $\l^2$-norm importance sampling, assemble chosen cols into $q\times q$ matrix $W$ with rescaling.

      \smallskip
      \State \textbf{(SVD and Filter)}
      \State Computer the top $k$ left singular vectors of $W$ $\v u_1, \v u_2,...,\v u_k$ that correspond to singular values $\sig_1,\sig_2,...,\sig_k$ larger than $\sig$.
      \State Convert $\v u_\l\in \mathbb R^q$ into vectors $\v v_\l\in \mathbb R^n$ by
      \[
        \v v_k = \frac{S^\dag\v u_k}{|W^\dag \v u_k|}.
      \]

      \State Output $\v v_1, \v v_2,...,\v v_k$(the low rank approximation to $T$ is $\td T_{\le k}=T\sum_{\l=1}^k \v v_\l\v v_\l^\dag$).
    \end{algorithmic}
  \end{algorithm}

  $\l^2$-norm importance sampling was introduced and defined in Sec. \ref{sec:intro_qc}.

  The time complexity of the quantum-inspired classical algorithm is $\td\O\left(\log(mn)\cdot\frac{|T|^{24}}{\sig_k^{24}\eps^{12}\kappa^{6}}\right)$, where $\td\O$ hides the log factors incurred by amplifying the failure probability\footnote{Note that $\eps$ and $\kappa$ here are error parameters, not to confuse them with the DP parameters.}. Focusing on $m,n,k$ (ignoring error parameters) and using the same techniques as before, we get a time complexity $\td\O(k^{12}\cdot\log(mn))$, a large slowdown versus the quantum algorithm in terms of the exponent of $k$. However, for the quantum algorithm the exponent of $\log(mn)$ depends on the quantum RAM model and remains unclear for the time being, so we cannot give a full comparison between the quantum and classical algorithm in terms of the time complexity.

  Note that both algorithms are rather complicated, so it is recommended to refer to the original literature \cite{q_recommend_sys,ewin_tang_ciq} for complete details if you are still not clear about some aspects of either algorithms.





\end{document}